\documentclass[oldversion]{aa}  
\usepackage{graphicx,amsmath}
\usepackage{amssymb}
\usepackage{color}
\usepackage{natbib}             
\usepackage{url}
\usepackage{multirow}
\usepackage{threeparttable}
\bibpunct{(}{)}{;}{a}{}{,} 

\def\approxinf{%
  \def\p{%
    \setbox0=\vbox{\hbox{$<$}}%
    \ht0=0.6ex \box0 }%
  \def\s{%
    \vbox{\hbox{$\sim$}}%
  }%
  \mathrel{\raisebox{0.7ex}{%
      \mbox{$\underset{\s}{\p}$}%
    }}%
}

\begin{document}

   \title{No hydrogen exosphere detected around the super-Earth HD\,97658 b}

   \author{
   V.~Bourrier\inst{1},
   D.~Ehrenreich\inst{1}, 
   G.~King\inst{3}   
   A.~Lecavelier des Etangs\inst{2},
   P.J. Wheatley\inst{3}
   A.~Vidal-Madjar\inst{2},  
   F.~Pepe\inst{1},\and      
   S.~Udry\inst{1}}

\authorrunning{V.~Bourrier et al.}
\titlerunning{Lyman-$\alpha$ observations of the super-Earth HD\,97658 b}
\offprints{V.B. (\email{vincent.bourrier@unige.ch})}

\institute{
Observatoire de l'Universit\'e de Gen\`eve, 51 chemin des Maillettes, 1290 Sauverny, Switzerland
\and 
Institut d'astrophysique de Paris, UMR7095 CNRS, Universit\'e Pierre \& Marie Curie, 98bis boulevard Arago, 75014 Paris, France 
\and   
Department of Physics,University of Warwick, Coventry CV4 7AL, UK
}
   
   \date{} 

  \abstract
{The exoplanet HD\,97658b provides a rare opportunity to probe the atmospheric composition and evolution of moderately irradiated super-Earths. It transits a bright K star at a moderate orbital distance of 0.08\,au. Its low density is compatible with a massive steam envelope that could photodissociate at high altitudes and become observable as escaping neutral hydrogen. Our analysis of three transits with HST/STIS at Lyman-$\alpha$ reveals no such signature, suggesting that the thermosphere of HD\,97658b is not hydrodynamically expanding and is subjected to a low escape of neutral hydrogen ($<$10$^{8}$\,g\,s$^{-1}$ at 3$\sigma$).\\
Using HST/STIS Lyman-$\alpha$ observations and Chandra/ACIS-S \& XMM-Newton/EPIC X-ray observations at different epochs, we find that HD\,97658 is in fact a weak and soft X-ray source with signs of chromospheric variability in the Lyman-$\alpha$ line core. We determine an average reference for the intrinsic Lyman-$\alpha$ line and X-EUV (XUV) spectrum of the star, and show that HD\,97658 b is in mild conditions of irradiation compared to other known evaporating exoplanets with an XUV irradiation about three times lower than the evaporating warm Neptune GJ436 b. This could be the reason why the thermosphere of HD\,97658b is not expanding: the low XUV irradiation prevents an efficient photodissociation of any putative steam envelope. Alternatively, it could be linked to a low hydrogen content or inefficient conversion of the stellar energy input. The HD\,97658 system provides clues for understanding the stability of low-mass planet atmospheres in terms of composition, planetary density, and irradiation. \\
Our study of HD\,97658 b can be seen as a control experiment of our methodology, confirming that it does not bias detections of atmospheric escape and underlining its strength and reliability. Our results show that stellar activity can be efficiently discriminated from absorption signatures by a transiting exospheric cloud. They also highlight the potential of observing the upper atmosphere of small transiting planets to probe their physical and chemical properties.}


 \maketitle

\section{Introduction}

A substantial fraction of known exoplanets orbit extremely close to their stars, within a tenth of an astronomical unit. This population displays a wide diversity in nature, from gaseous giants to super-Earths and even disintegrating rocky cores, raising many questions about the formation and evolution of such objects and their possible relations. The intense X-ray and extreme ultraviolet irradiation from the parent star can lead a hydrogen-rich thermosphere to lose its stability, expand hydrodynamically, and evaporate (\citealt{VM2003}).\\

An extended atmosphere produces a much deeper absorption than the lower atmospheric layers when observed in the spectral lines of elements that are abundant at high altitudes and/or associated with strong electronic transitions. Transit observations in the Lyman-$\alpha$ line of their host star first revealed the existence of extended atmospheres of neutral hydrogen around the hot Jupiters HD\,209458b (\citealt{VM2003,VM2004}; \citealt{BJ2007,BJ2008}; \citealt{VM2008}; \citealt{Ehrenreich2008}; \citealt{BJ_Hosseini2010}) and HD\,189733b (\citealt{Lecav2010,Lecav2012}; \citealt{Bourrier2013}). Observations of other lines in the far UV allowed the detection of heavy metals and ions carried to high altitudes by collisions with the hydrogen outflow (e.g., oxygen, carbon, and magnesium; \citealt{VM2004},  \citealt{Linsky2010}, \citealt{VM2013}, \citealt{BJ_ballester2013}, \citealt{Fossati2010}; \citealt{Haswell2012}), confirming that the upper atmosphere of these giant planets is in a state of blow-off.\\

Many questions remain to be answered about the effect of irradiation on the formation and the structure of extended atmospheres. The strong energy input into hot Jupiters is the source for their evaporation (e.g., \citealt{Lammer2003}, \citealt{Lecav2004}; \citealt{Yelle2004,Yelle2006}; \citealt{GarciaMunoz2007}; \citealt{Owen2012}, \citealt{Koskinen2013a,Koskinen2013b}, \citealt{Johnstone2015}), but the detection of a Lyman-$\alpha$ absorption from the partially transiting upper atmosphere of the warm Jupiter 55 Cnc b first hinted that milder conditions of irradiation can lead to atmospheric expansion (\citealt{Ehrenreich2012}). Interestingly, 55 Cnc b is close to the orbital distance limit where the atmosphere of a giant planet is expected to lose its stability (\citealt{Koskinen2007}). The moderate depth of 55 Cnc b absorption signature (\citealt{Ehrenreich2012}), in comparison to HD\,209458b and HD\,189733b, can be explained if its exosphere fills about a third of the Roche lobe, consistent with the lower irradiation of the planet. \\

Recently, transit observations of the warm Neptune GJ\,436 b in the Lyman-$\alpha$ line at three different epochs revealed deep, repeatable exospheric transits reaching up to 60\%, and lasting for much longer than the optical transit (\citealt{Kulow2014}, \citealt{Ehrenreich2015}). ``Radiative braking'' (\citealt{Bourrier2015}) and interactions with the wind of its M dwarf host star (\citealt{Bourrier2016}) were shown to shape the neutral hydrogen exosphere of GJ\,436b into a giant coma that surrounds the planet and trails for millions of kilometers behind it. Counterintuitively, GJ\,436 b observations revealed that much larger atmospheric signals could be retrieved from the upper atmosphere of moderately irradiated, low-mass planets than from more heavily irradiated massive planets. This opens thrilling perspectives for the characterization of the many small planets in the sub-Neptune and super-Earth mass regimes, which have no equivalent in the solar system and whose nature and evolution remain mysterious. Probing the upper extended atmosphere of such planets will be particularly important because their lower atmosphere, less extended and with a heavier composition, will be more difficult to characterize. The non-detection of a neutral hydrogen exosphere around the super-Earth 55 Cnc e (\citealt{Ehrenreich2012}) hinted at the presence of a high-weight atmosphere -- or the absence of an atmosphere -- recently supported by the study of its brightness map in the IR (\citealt{Demory2016}). Studying the atmospheric escape of small planets will not only bring insights into their nature, but will also help understand their formation and evolution. \\

The exoplanet HD\,97658 b ($M_{p}$ = 7.9$\pm$0.7\,M$_\oplus$; $R_{p}$ = 2.39\,R$_\oplus$) was discovered by a radial-velocity survey (\citealt{Howard2011}), and was later detected in transit across its 9.7$\pm$2.8\,Gy old parent star (\citealt{Dragomir2013_HD976}; \citealt{Bonfanti2016}). Measurement of the planetary radius at 4.5\,$\mu$m with Spitzer have allowed  the planet bulk density to be refined to $\rho$ = 3.9$\stackrel{+0.7}{_{-0.6}}$\,g\,cm$^{-3}$ (\citealt{Vangrootel2014}). While its exact nature remains to be unveiled, \citet{Vangrootel2014} have predicted that HD\,97658 should have a large rocky core accounting for up to 60\% of the planet mass, an envelope of water and ices accounting for up to 40\% of the planet mass, and less than 2\% of hydrogen and helium. If its composition is indeed water-rich, HD\,97658 orbits close enough to its K1-dwarf star ($P$ = 9.5\,days, T$_{eq}\sim$725\,K; \citealt{Knutson2014}) that its temperature should allow for the formation of a thick envelope of steam. The brightness of HD\,97658 b host star (V=7.7, $d$=21.1\,pc) makes it one of the handful of small planets currently amenable to atmospheric characterization. However, given the shallowness of the planetary transit seen at optical or near-infrared (NIR) wavelengths ($\sim$0.09\%) and the small expected scale height of a steam envelope, its detection would be extremely challenging through infrared absorption. As it is, transmission spectroscopy of HD\,97658b performed in the near IR with WFC3 (\citealt{Knutson2014}) is consistent with flat transmission spectrum models, indicating either a cloudy atmosphere or a cloudless, water-rich atmosphere. Consequently, HD\,97658 b is a very good target for transit observations of its upper atmosphere. Photodissociation of a steam envelope could produce OH and H at high altitudes (\citealt{Wu1993}; \citealt{Jura2004}). Here, we present a search for neutral hydrogen escape through dedicated Lyman-$\alpha$ line observations. These observations are complemented by X-ray measurements to estimate the high-energy stellar irradiation. This study is comparable to the one performed for the hot Jupiter HD\,189733b, which transits a star with the same type as HD\,97658 and was shown to be evaporating (\citealt{Lecav2010,Lecav2012}, \citealt{Bourrier2013}). Differences in the age and coronal activity of the host stars, in addition to the different orbital and bulk properties of their planets, may induce very different evaporation regimes.\\
Lyman-$\alpha$ observations of HD\,97658 at different epochs are analyzed and interpreted in Sect.~\ref{sec:data ana}. They are used in Sect.~\ref{sec:recons} to estimate the intrinsic Lyman-$\alpha$ line of HD\,97658 and the properties of the interstellar medium (ISM) in its direction. We use these measurements, along with that of the stellar X-ray emission, to study the XEUV irradiation of HD\,97658 b in Sect.~\ref{sect:X-EUV}. In Sect.~\ref{sec:EVE_sim} we compare the Lyman-$\alpha$ observations with numerical simulations of the planet transit with the EVaporating Exoplanet (EVE) code to constrain the presence and properties of a putative extended thermosphere and exosphere of neutral hydrogen. We discuss our results and their importance for the nature of HD\,97658b in Sect.~\ref{sec:discuss}. The properties of the system relevant to our study are given in Table~\ref{tab:param_sys}.\\

\begin{table}[tbh]
\caption{Physical parameters for the HD\,97658 system.}                                                 
\begin{tabular}{llcccc}
\hline
\hline
\noalign{\smallskip}
Parameters                                        & Symbol      & Value          \\
\noalign{\smallskip}
\hline
\noalign{\smallskip}
Distance from Earth             & $D_{\mathrm{*}}$               &    21.11\,pc    \\
\noalign{\smallskip}
Star radius                             & $R_{\mathrm{*}}$       &    0.74$\,R_{\mathrm{\sun}}$   \\
\noalign{\smallskip}
Star mass                                       & $M_{\mathrm{*}}$      &    0.77$\,M_{\mathrm{\sun}}$    \\
\noalign{\smallskip}
Planet radius           & $R_{\mathrm{p}}$                       &    2.39$\,R_{\mathrm{Earth}}$         \\\noalign{\smallskip}
Planet mass                     & $M_{\mathrm{p}}$               &    7.55$\,M_{\mathrm{Earth}}$   \\
\noalign{\smallskip}
Orbital period                          & $P_{\mathrm{p}}$               &    9.49$\,days$  \\
\noalign{\smallskip}
Transit center                          & $T_{\mathrm{0}}$                       &   2456665.46415$\,BJD$  \\
\noalign{\smallskip}
Semi-major axis                         & $a_{\mathrm{p}}$                       &    0.08$\,au$ \\
\noalign{\smallskip}
Eccentricity                            & $e$                                            &    0.078      \\
\noalign{\smallskip}
Argument of periastron  & $\omega$                                                       &    71$^{\circ}$       \\
\noalign{\smallskip}
Inclination                       & $i_{\mathrm{p}}$                             &    89.14$^{\circ}$    \\
\noalign{\smallskip}
\hline
\hline
\multicolumn{6}{l}{Note: All parameters come from \citealt{Vangrootel2014}}\\
\multicolumn{6}{l}{except for $T_{\mathrm{0}}$, $P_{\mathrm{p}}$, and $R_{\mathrm{p}}$ from \citealt{Knutson2014}}\\
\end{tabular}
\label{tab:param_sys}
\end{table}


\section{Analysis of the resolved Lyman-$\alpha$ line}
\label{sec:data ana}

\subsection{Transit observations}

\begin{figure}     
\includegraphics[trim=2cm 4.5cm 1.5cm 5.7cm,clip=true,width=\columnwidth]{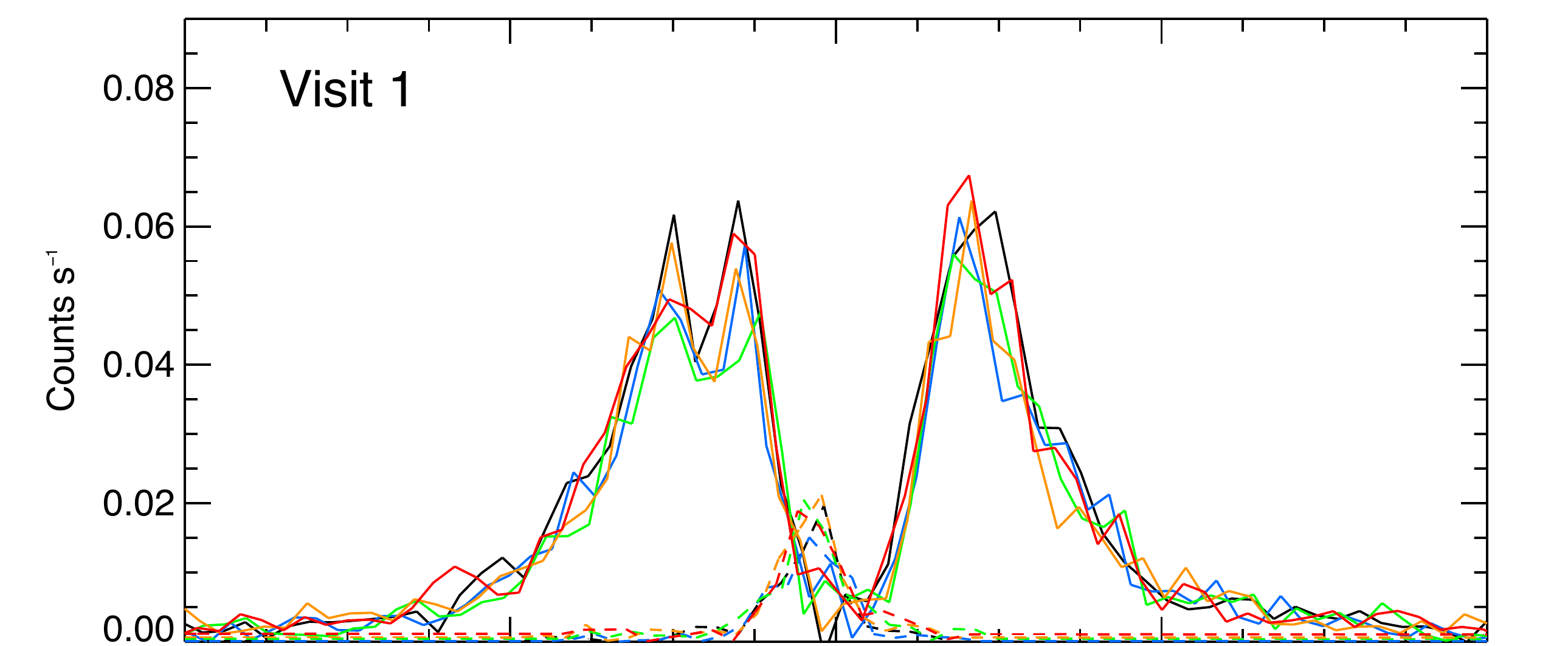}
\includegraphics[trim=2cm 4.5cm 1.5cm 5.7cm,clip=true,width=\columnwidth]{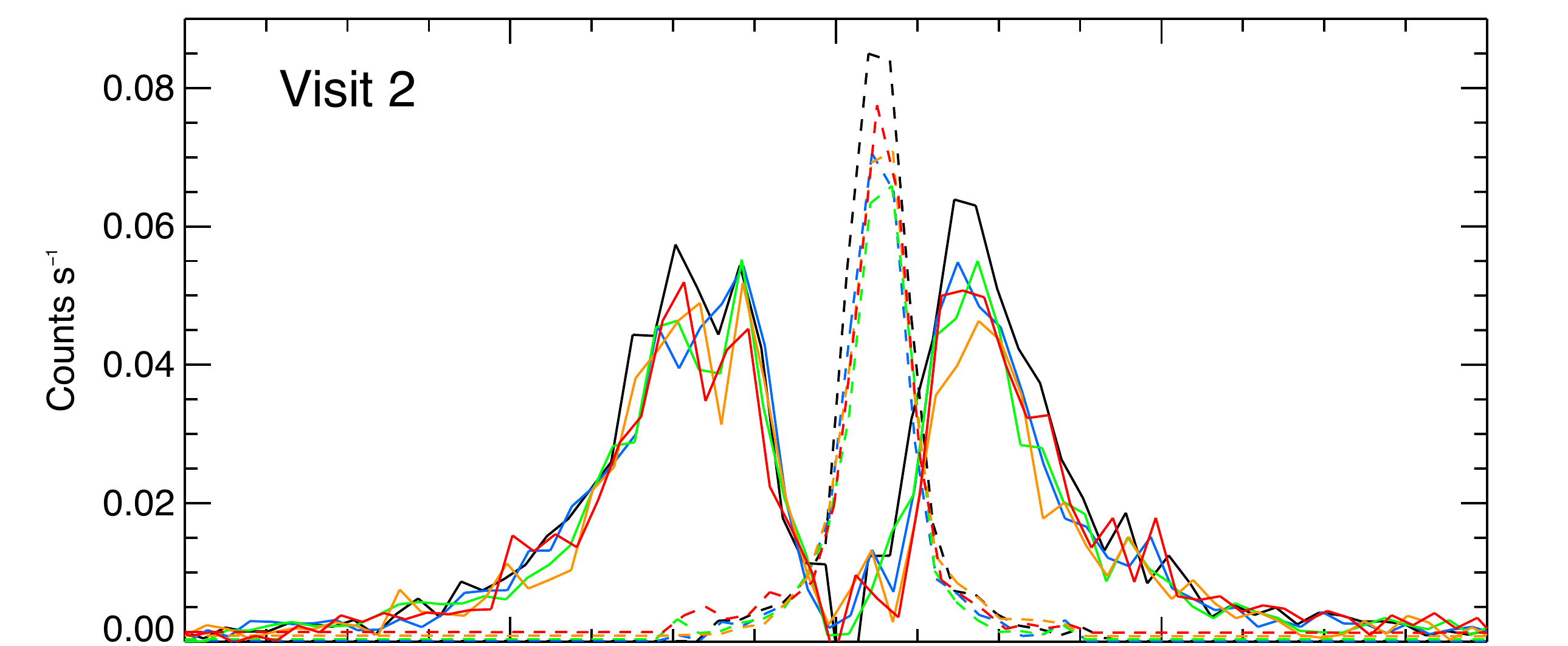}
\includegraphics[trim=2cm 2.8cm 1.5cm 5.7cm,clip=true,width=\columnwidth]{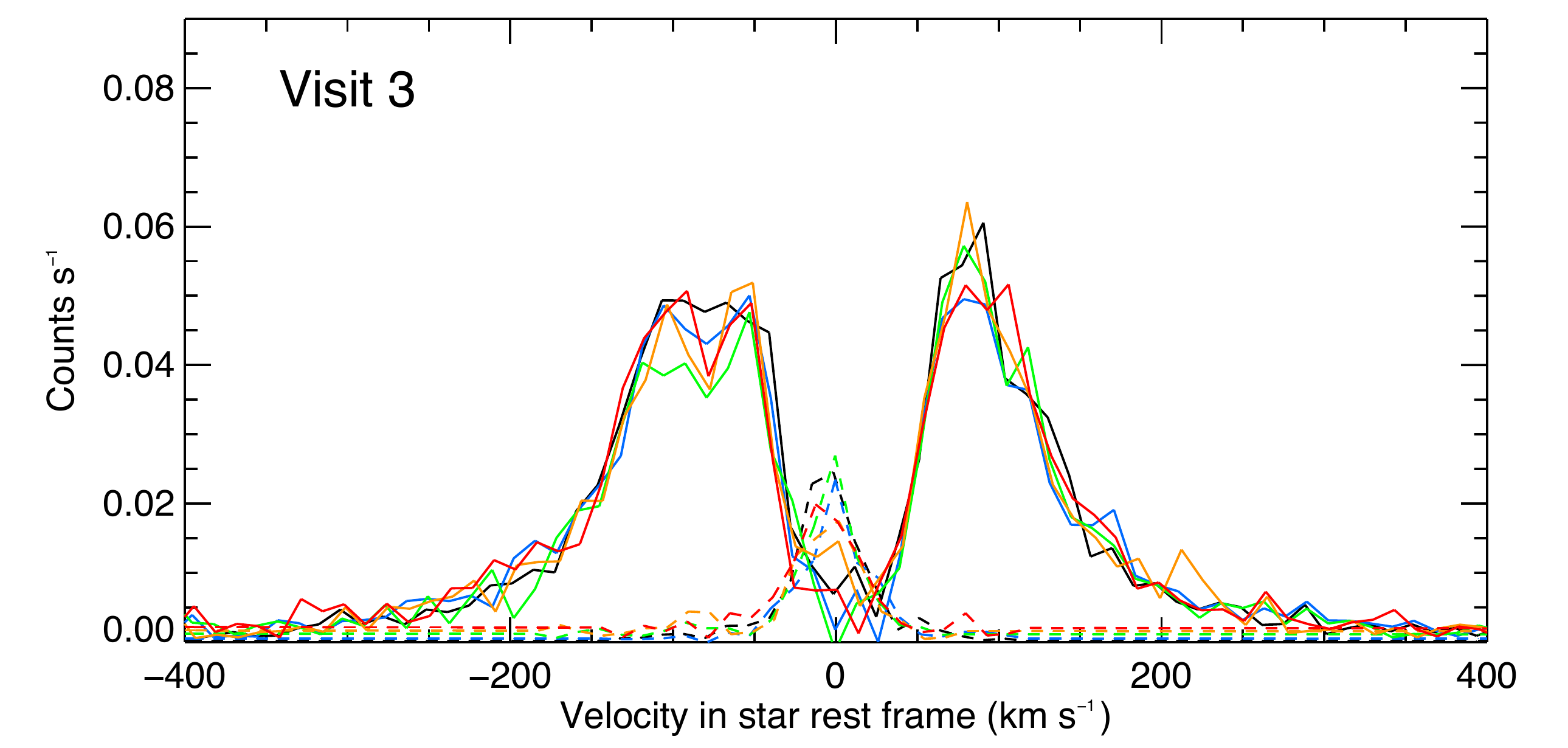}
\caption[]{Raw spectra of the stellar Lyman-$\alpha$ line (solid line) after correction from the geocoronal emission line (superimposed as a dashed line). Visits 1, 2, and 3 are displayed from top to bottom, and in each panel the colors correspond to the  HST orbit at consecutive orbital phases (increasing from black, blue, green, orange, red).}
\label{fig:airglow}
\end{figure}

We observed HD\,97658 in the H\,{\sc i} Lyman-$\alpha$ line (1215.6702\,\AA) with the Space Telescope Imaging Spectrograph (STIS) instrument on board the Hubble Space Telescope (HST). Three visits of five HST orbits each were obtained in the frame of the GO Program 13820 (PI: Ehrenreich), scheduled so that Visit 1 (April 2015) covered one transit of HD\,97658b, Visit 2 (December 2015) covered the transit and post-transit phase, and Visit 3 (March 2016) covered the pre-transit and transit phase (Table~\ref{obs_log}). The time-tagged data obtained with the G140M grating were reduced with the CALSTIS pipeline. Each HST orbit was divided into $\sim$300\,s exposures, as a compromise between signal-to-noise ratio (S/N) and temporal resolution. In all visits, this yielded six subexposures for the first orbit (the target acquisition reduces the duration of its scientific exposition) and seven for the other orbits. In the raw data (Fig.~\ref{fig:airglow}), the stellar Lyman-$\alpha$ emission line is superimposed with the geocoronal airglow emission from the upper atmosphere of the Earth (\citealt{VM2003}). The airglow contamination was limited by the use of STIS narrow slit of 52''$\times$0.05'', and can be  estimated and removed from the final 1D spectra using CALSTIS. It is nonetheless recommended to treat with caution the regions where the airglow is stronger than the stellar flux. The strength and position of the airglow varies with the epoch of observation, and in a first step we excluded from our analysis the spectral ranges [-27 ; 12]\,km\,s$^{-1}$ (Visit 1), [-13 ; 40]\,km\,s$^{-1}$ (Visit 2), and [-21 ; 31]\,km\,s$^{-1}$ (Visit 3), defined in the stellar rest frame. We note that correcting manually for the airglow, using different areas of the 2D images to build its profile, did not change the final spectra significantly. For all visits, we also limited the analysis to the spectral range where the S/N is high enough, from -305 to 305\,km\,s$^{-1}$. \\

\begin{table*}
\centering
\begin{tabular}{lccccc}
\hline
\hline
\noalign{\smallskip}    
                   & 12 February 2015    & 8 April 2015       & 4 June 2015 & 11 December 2015 & 5 March 2016 \\
\hline
\noalign{\smallskip}
Lyman-$\alpha$ & HST/STIS Visit 0 & HST/STIS Visit 1 &     -     & HST/STIS Visit 2 & HST/STIS Visit 3 \\ 
\noalign{\smallskip}
X-ray              &       -         &         -         & XMM-Newton/EPIC & Chandra/ACIS-S & Chandra/ACIS-S \\  
\noalign{\smallskip}
\hline
\hline
\end{tabular}
\caption{Log of HD\,97658b observations. In the last two epochs, the Lyman-$\alpha$ and X-ray observations are contemporaneous.}
\label{obs_log}
\end{table*}

\subsection{Systematic variations.}

\begin{figure}     
\includegraphics[trim=3cm 4.6cm 7cm 8cm,clip=true,width=\columnwidth]{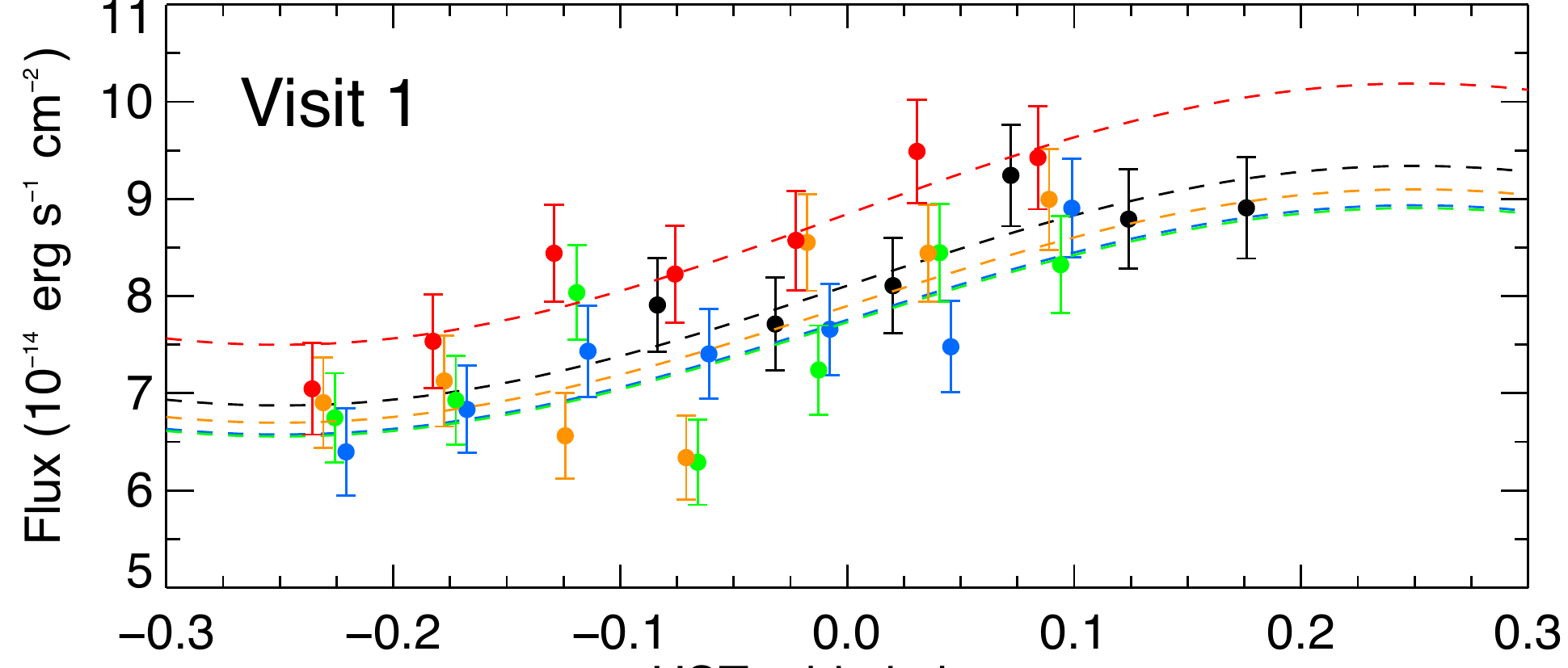}
\includegraphics[trim=3cm 4.6cm 7cm 8cm,clip=true,width=\columnwidth]{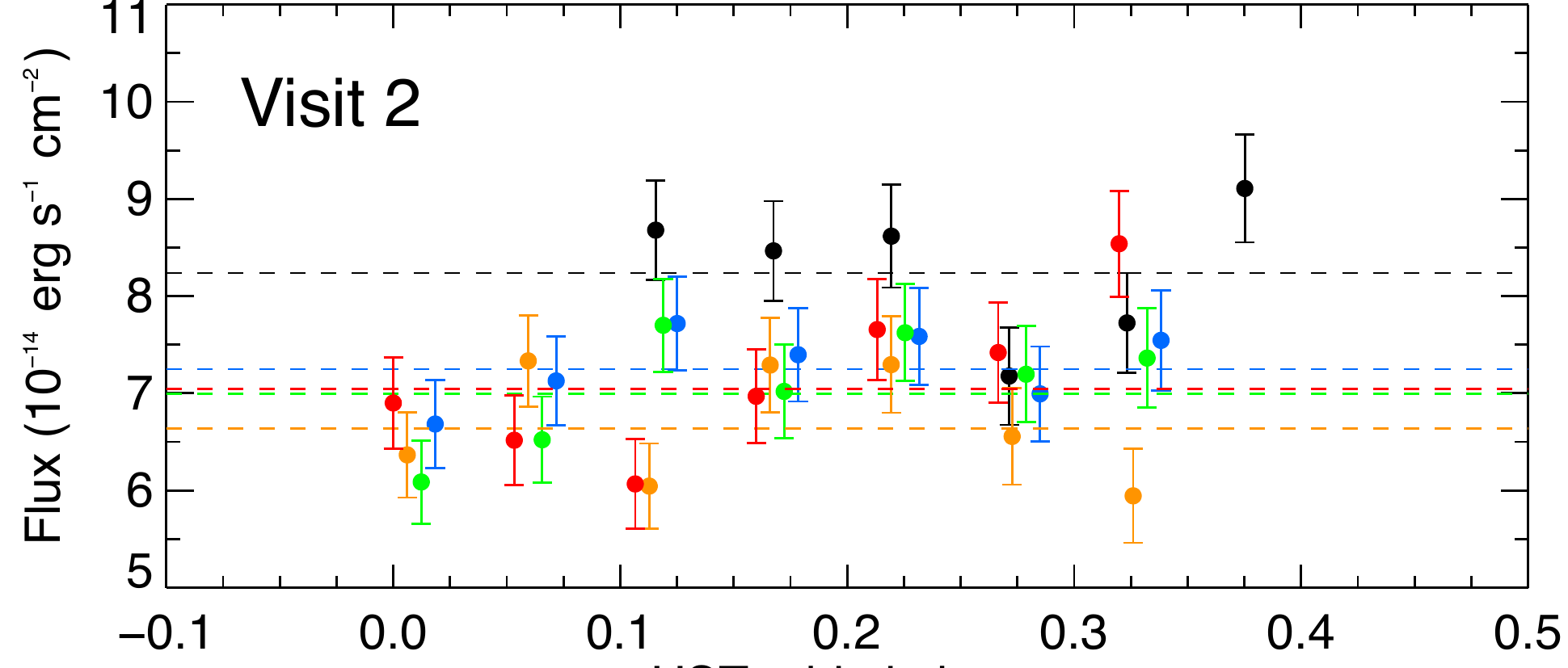}
\includegraphics[trim=3cm 4cm 7cm 8cm,clip=true,width=\columnwidth]{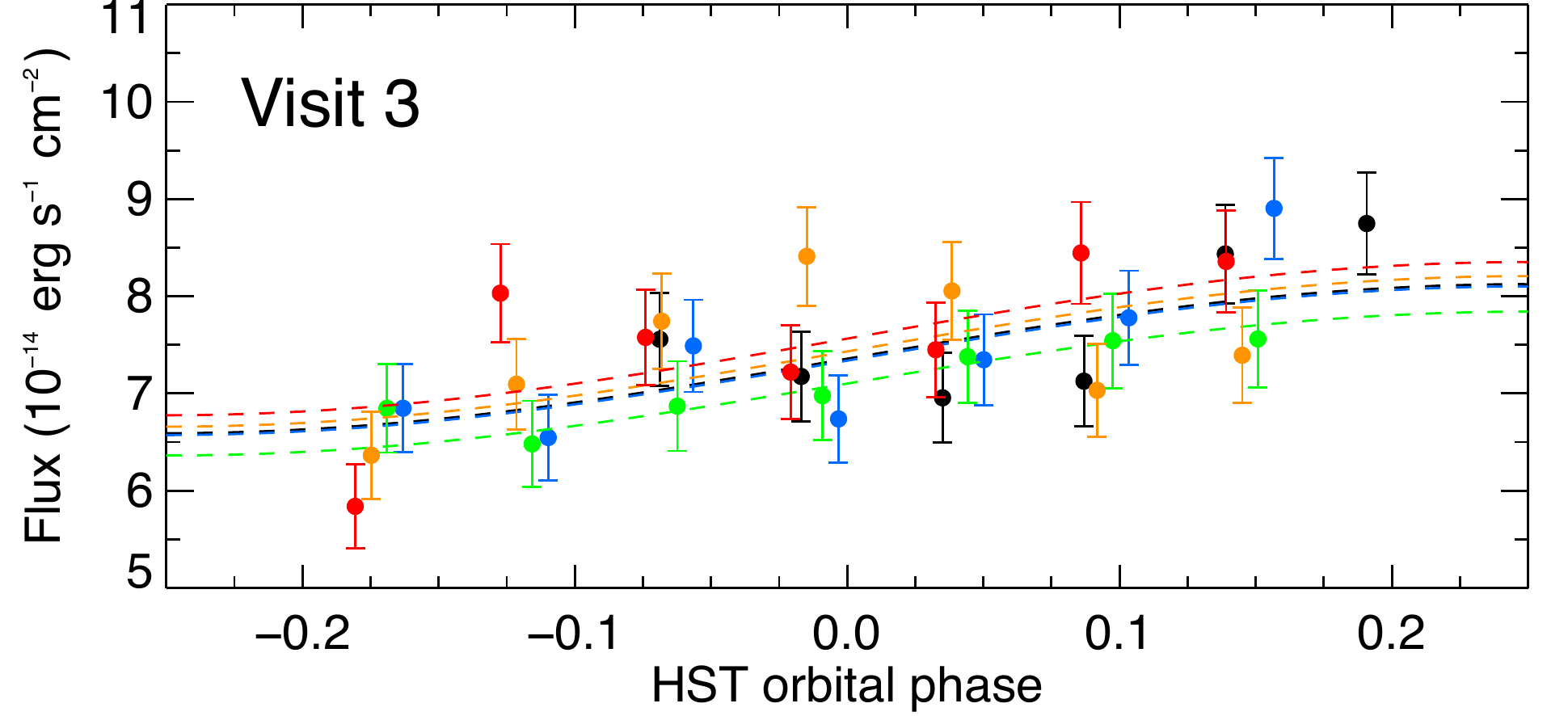}
\caption[]{Lyman-$\alpha$ fluxes for 300\,s exposures integrated over the entire line and phase-folded on the HST orbital period. The dashed lines correspond to the best fit for the $F_{\mathrm{corr}}$ function. Since the contribution of the breathing effect $f_{\mathrm{breath}}$  to this fit is the same at a given HST phase, the best estimation for the intrinsic stellar flux in each orbit $F_{\mathrm{nom}}$ can be read at phase = 0 (see Eq.~1). The color code is the same as in Fig.~\ref{fig:airglow}, and visits 1, 2, and 3 are displayed from top to bottom.}
\label{fig:breath_flux}
\end{figure}

\begin{figure}     
\includegraphics[trim=3cm 4.6cm 7cm 8cm,clip=true,width=\columnwidth]{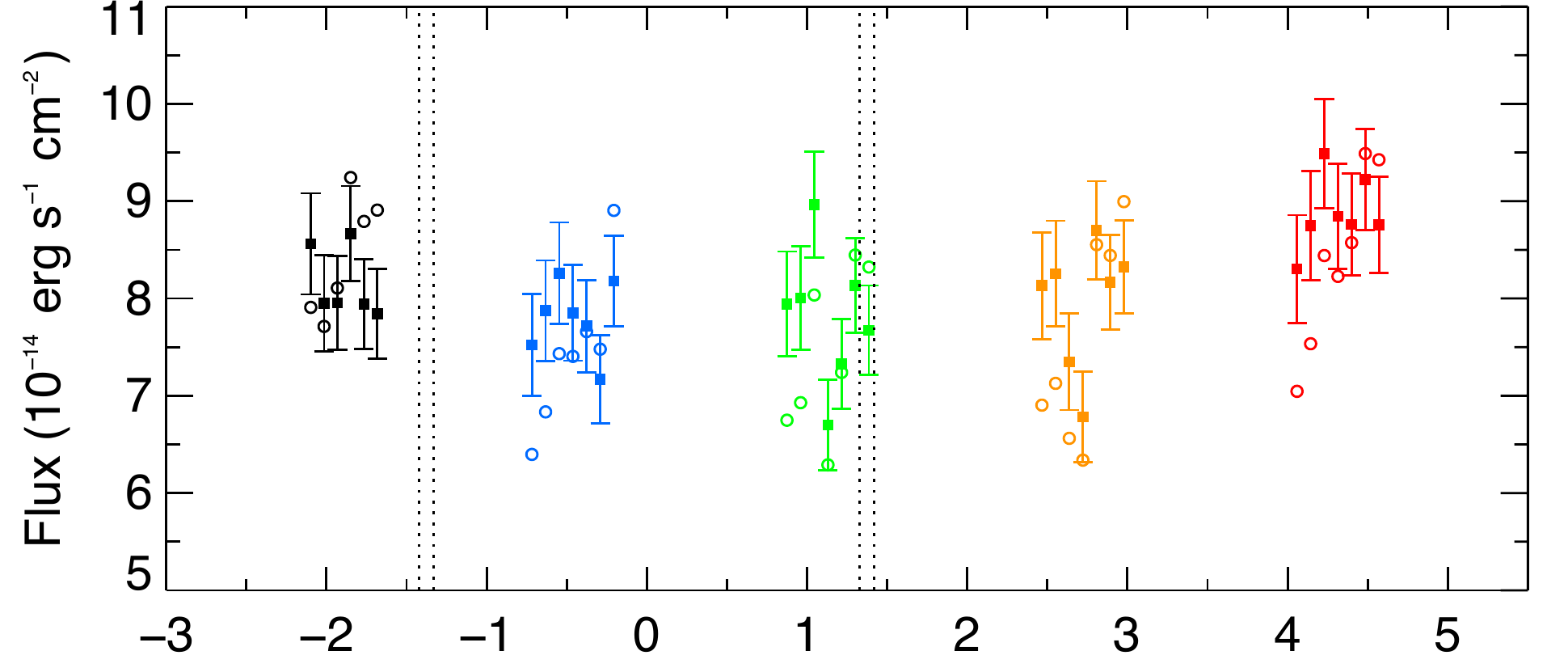}
\includegraphics[trim=3cm 4.6cm 7cm 8cm,clip=true,width=\columnwidth]{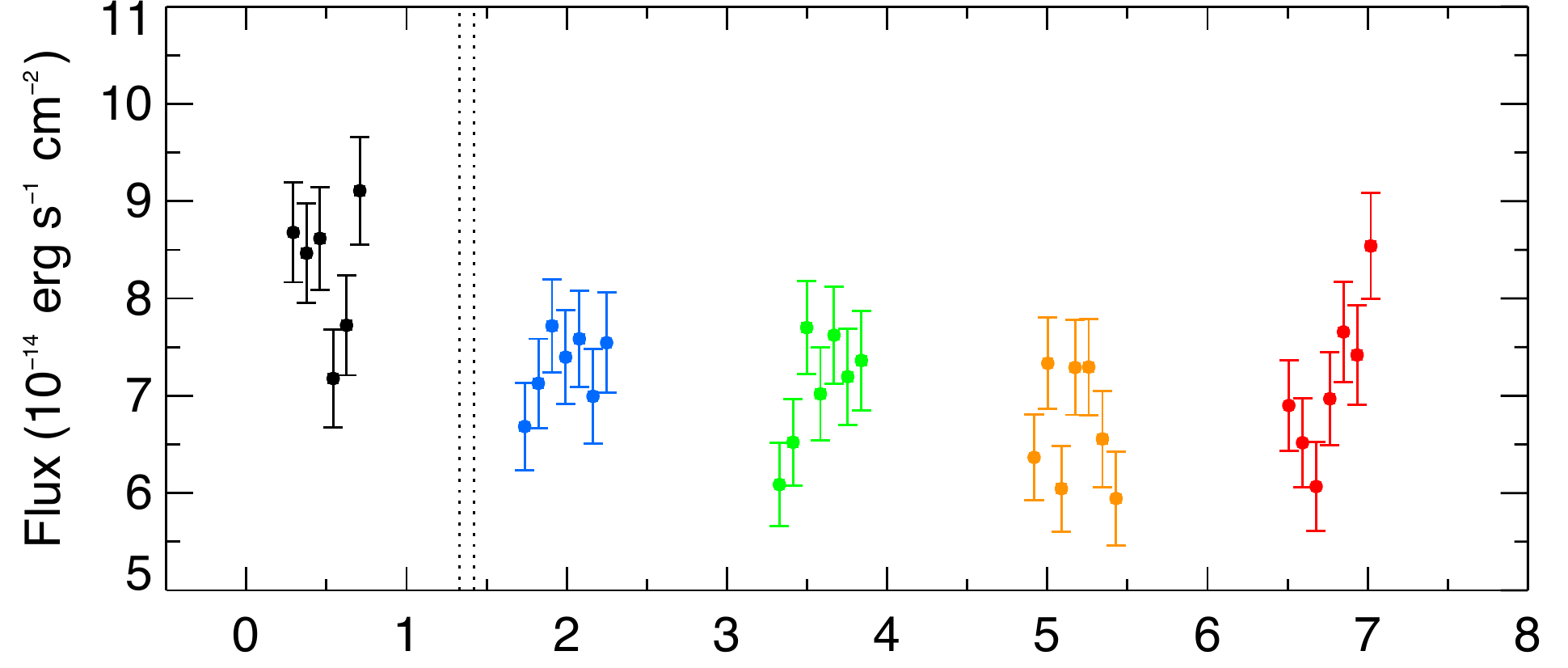}
\includegraphics[trim=3cm 4cm 7cm 8cm,clip=true,width=\columnwidth]{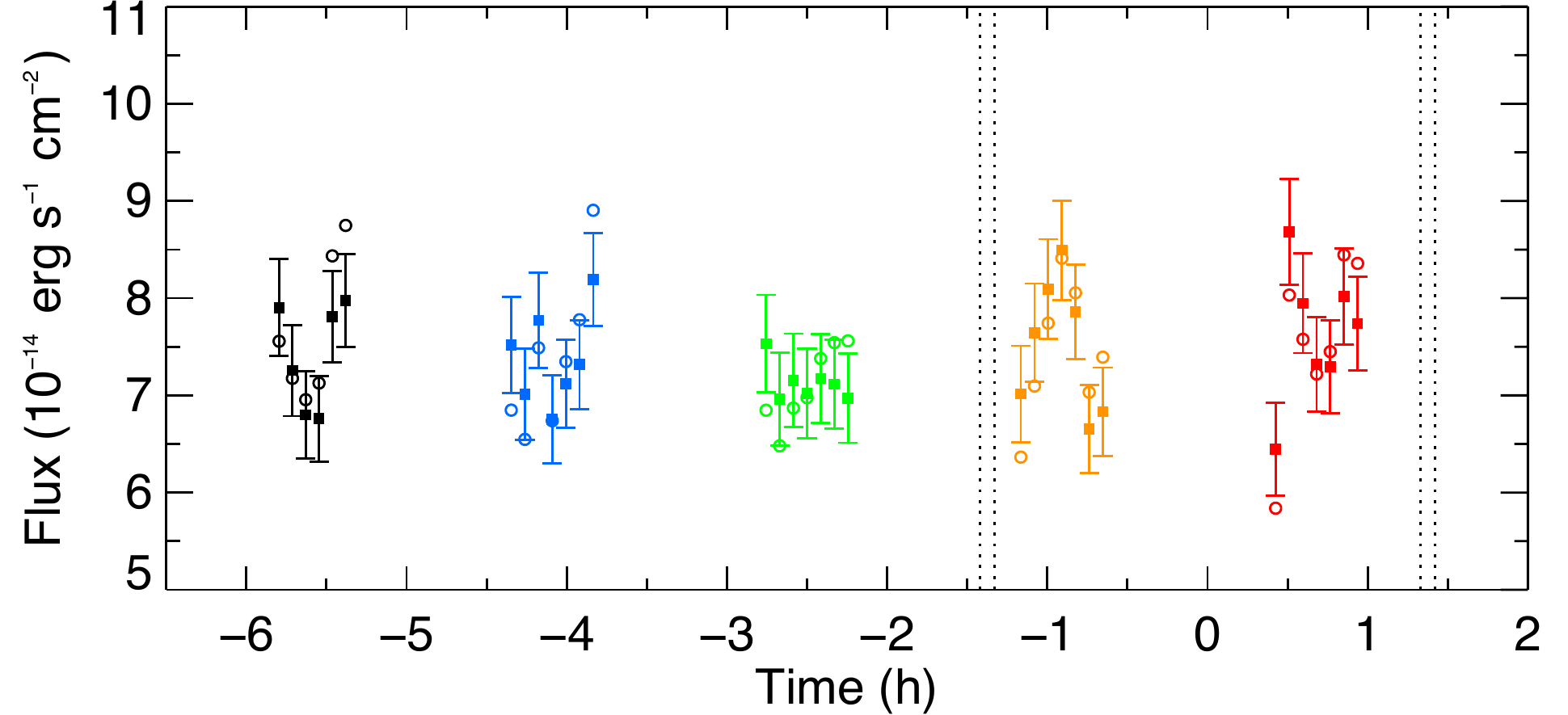}
\caption[]{Raw fluxes uncorrected for the breathing effect (empty disks), as a function of time, relative to the center of the planetary transit. After correction by the $f_{\mathrm{breath}}$ function in Visits 1 and 3 (filled squares), the fluxes within a given orbit are,  on average, equal to the best-fit value obtained for $F_{\mathrm{nom}}$ in this orbit. Each point stands for a 300 s exposure spectrum summed over the whole Lyman-$\alpha$ line. Vertical dotted lines show the beginning and end of ingress and egress of the transit. The color code is the same as in Fig.~\ref{fig:airglow}, and visits 1, 2, and 3 are displayed from top to bottom.}
\label{fig:corr_flux}
\end{figure}

We checked for variations on short timescales within a given HST orbit caused by the telescope ``breathing''. This effect is known to affect STIS observations in the UV and in the visible because of variations of the telescope throughput caused by the changes in temperature that the HST experiences during each orbit (e.g., \citealt{Bourrier2013}). The shape and amplitude of the breathing variations can change between visits of the same target, but the orbit-to-orbit variations within a single visit are both stable and highly repeatable (e.g. \citealt{Brown2001}; \citealt{Sing2008a}; \citealt{Huitson2012}; \citealt{Ehrenreich2015}). We thus modeled the breathing effect using a Fourier series decomposition with a period equal to that of the HST around the Earth ($P_{\mathrm{HST}}$ = 96\,min), and adjusted the following function to the 300\,s exposure spectra in each visit,
\begin{align}
F_{\mathrm{corr}}(t) &= F_{\mathrm{nom}}(k_{orb}) \times f_{\mathrm{breath}}(\phi_{\mathrm{HST}}(t)) \\  \nonumber
f_{\mathrm{breath}}(\phi_{\mathrm{HST}})&= 1+\sum_{i=0}^{n} a_{\mathrm{i}}\,sin( 2\,i\,\pi\,\phi_{\mathrm{HST}} ) \\ \nonumber
\phi_{\mathrm{HST}}(t)&= \frac{t-t_{\mathrm{ref}}}{P_{\mathrm{HST}}},  \nonumber
\label{eq:corr_func}
\end{align}
with $\phi_{\mathrm{HST}}(t)$ the HST orbital phase at a given absolute time $t$ and $f_{\mathrm{breath}}$ the breathing model function. Its degree $n$ and coefficients $a_{\mathrm{i}}$ are free parameters of the fit. The function $F_{\mathrm{nom}}$ corresponds to the nominal stellar flux unaffected by the breathing effect. Its value for each orbit, indexed by $k_{orb}$, is a free parameter of the model. This accounts for variations in the flux from one orbit to another arising for example from intrinsic stellar variability. We note that we tried defining $F_{\mathrm{nom}}$ as a polynomial function with its degree as a free parameter; there was no significant change in the results of the fit.\\
The breathing effect is achromatic, but it must be analyzed in wavelength bands where no spectral variations are found as a function of time that could be caused by the planetary atmosphere, which would risk being overcorrected. Typical exospheric signatures are located in the blue wing of the line, and the red wing can usually be used as a stable reference to characterize systematic variations. However, in the present case we found no spectro-temporal signatures that could clearly be attributed to the planet (this was validated {a posteriori} in Sect.~\ref{sec:line_variations}), and for each visit a thorough search for the breathing effect in all parts of the line yielded similar results for $f_{\mathrm{breath}}$. We thus fitted the $F_{\mathrm{corr}}$ function to the flux integrated in the entire Lyman-$\alpha$ line, using the Bayesian Information Criterion (BIC) as a merit function (\citealt{Crossfield2012}; \citealt{Cowan2012}).\\
Results are shown in Fig.~\ref{fig:breath_flux}. For Visits 1 and 3, we detected a significant breathing effect with order $n$=1, yielding respective BICs of 57 and 63. By comparison, the same fits performed with no breathing component ($n$=0) yielded BICs of 110 and 85. We detected no significant breathing variations in Visit 2: $n$=0 and 1 yielded similar BICS of 66, and $n>$2 yielded BICs$>$72. Variations of the stellar flux from one orbit to the other were derived from $F_{\mathrm{nom}}$ in Visits 1 and 2, and are discussed in Sect.~\ref{sect:shortterm_var}. We caution that the scientific exposure in the first HST orbit is shorter and begins at a later HST orbital phase than the other orbits (Fig.~\ref{fig:breath_flux}). Therefore, because the breathing-affected flux typically increases with HST phase (e.g., \citealt{Bourrier2013}; \citealt{Ehrenreich2012}; \citealt{Ehrenreich2015}), its average over the full duration of the first orbit can be overestimated compared to that of the other orbits. This underlines the importance of correcting for the breathing effect, which can only be done using data acquired in time-tag mode. Finally, we used the best fits for $F_{\mathrm{corr}}$ to correct each 300\,s exposure spectrum by the value of the breathing function $f_{\mathrm{breath}}$ at the time of mid-exposure (Fig.~\ref{fig:corr_flux}). \\

\subsection{Line profile variations}
\label{sec:line_variations}

A preliminary analysis of the breathing-corrected data revealed variations between the observed spectra, but none that could be attributed to the transit of an exosphere. Because of these variations, no exposure could be clearly identified as a reference for the unocculted, average, stellar Lyman-$\alpha$ line. Therefore, this reference was defined as the error-weighted mean of all exposures in a given visit and used to perform a more thorough analysis of the observations. \\

\subsubsection{Long-term evolution.}
\label{sect:longterm_var}

\begin{figure}     
\includegraphics[trim=2.5cm 5cm 5cm 2.5cm,clip=true,width=\columnwidth]{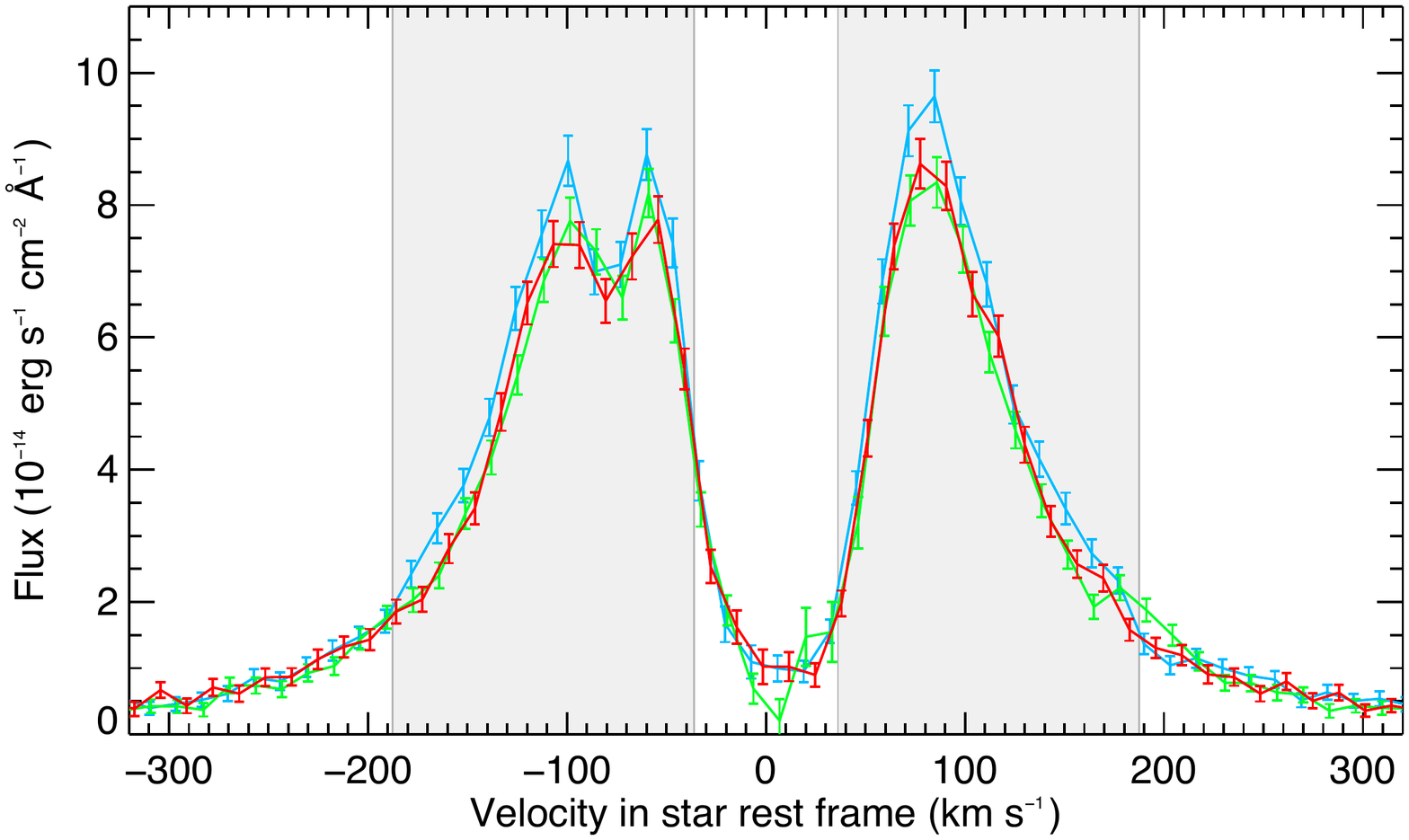}
\caption[]{Comparison between the reference spectra in Visits 1 (blue), 2 (green), and 3 (red). The spectra for Visits 2 and 3 display no significant differences except for a localized feature around 200\,km\,s$^{-1}$. They are also similar to the core and the far wings of the Visit 1 reference spectrum, which shows however a 11$\pm$2.7\% flux increase in the shaded gray regions.}
\label{fig:masters}
\end{figure}

First, we compared the reference spectra of the different visits to assess long-term variability in the shape of the Lyman-$\alpha$ line (Fig.\ref{fig:masters}). All visits show remarkably similar line profiles at high velocities in the wings of the line, with no variations at more than 2$\sigma$ beyond about $\pm$187\,km\,s$^{-1}$ (see also Fig.~\ref{fig:light_curves}). The only exception is a localized flux increase of 27$\pm$11\%, from about 170 to 215\,km\,s$^{-1}$, in Visit 2. The reference spectra are also similar in the core of the line (-36 to 36\,km\,s$^{-1}$). Although the Visit 2 spectrum displays spurious pixels that are likely biased by its higher airglow emission (Fig.~\ref{fig:airglow}), this gave us confidence that the background was properly corrected in the core of the Visits 1 and 3 spectra, and hereafter this range was included in our analysis of these visits. \\
Except for the core and the red wing signature, the reference spectra of Visits 2 and 3  are nearly identical. In contrast, the  Visit 1 spectrum shows a significant flux increase ($\sim$11.3$\pm$2.6\% at 4$\sigma$) in two wide bands roughly symmetrical in the stellar rest frame (36 $<$ $|$v$|$ $<$ 187\,km\,s$^{-1}$; Fig.~\ref{fig:masters}, Fig.~\ref{fig:light_curves}). These regions of higher flux at the peaks of the observed line likely trace the parts of the intrinsic line emitted by the lower transition region between the upper chromosphere of the star and its corona, that is at temperatures near 20~000\,K in the case of the quiet Sun (\citealt{Vernazza1981}). The flux increase in Visit 1 may thus indicate a higher level of chromospheric activity at this epoch. The similarity between the spectra of  Visits 2 and 3, and their difference with the Visit 1 spectrum, is also consistent with long-term variability of the star since Visit 2 was secured $\sim$8.1 month after Visit 1, while only $\sim$2.8 months separated Visits 2 and 3. We note that ISM absorption is highest  in the core of HD\,97658 Lyman-$\alpha$ line, and its intrinsic flux is thus lost from about -30 to 50\,km\,s$^{-1}$ (Sect.~\ref{sec:recons}; Fig.~\ref{fig:intrinsic_line}). The flux in the core of the {observed} line comes from the convolution of the ISM-absorbed spectrum farther from the core by STIS instrumental response, and while it still contains information that can be used to reconstruct the intrinsic line (Sect.~\ref{sec:recons}), we found that it is too blended to reflect directly the line variability between the epochs. Finally, the line wings beyond $187$\,km\,s$^{-1}$ are little affected by ISM absorption and directly trace the intrinsic line. The stability that they show over the three visits is consistent with their formation in the colder regions of the chromosphere.   \\

\subsubsection{Short-term variations.}
\label{sect:shortterm_var}

\begin{figure}     

\includegraphics[trim=2.5cm 2.9cm 7cm 10.3cm,clip=true,width=\columnwidth]{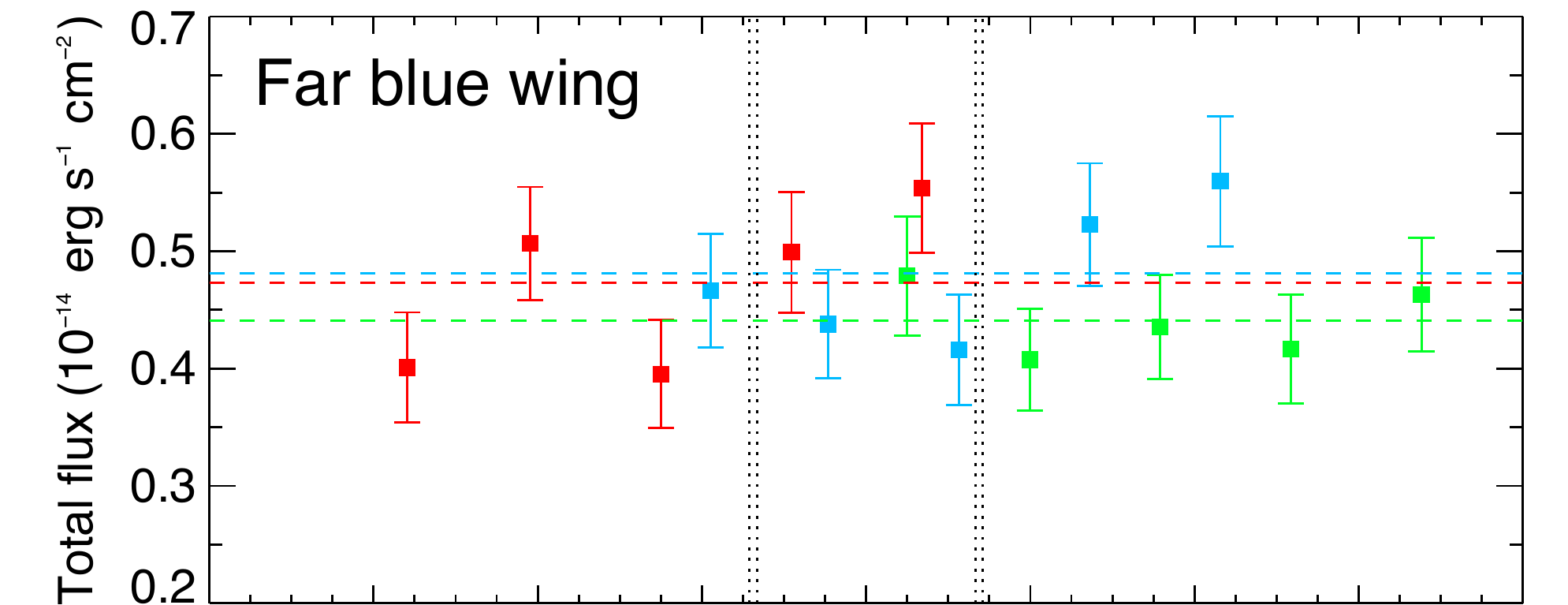}
\includegraphics[trim=2.5cm 2.9cm 7cm 10.3cm,clip=true,width=\columnwidth]{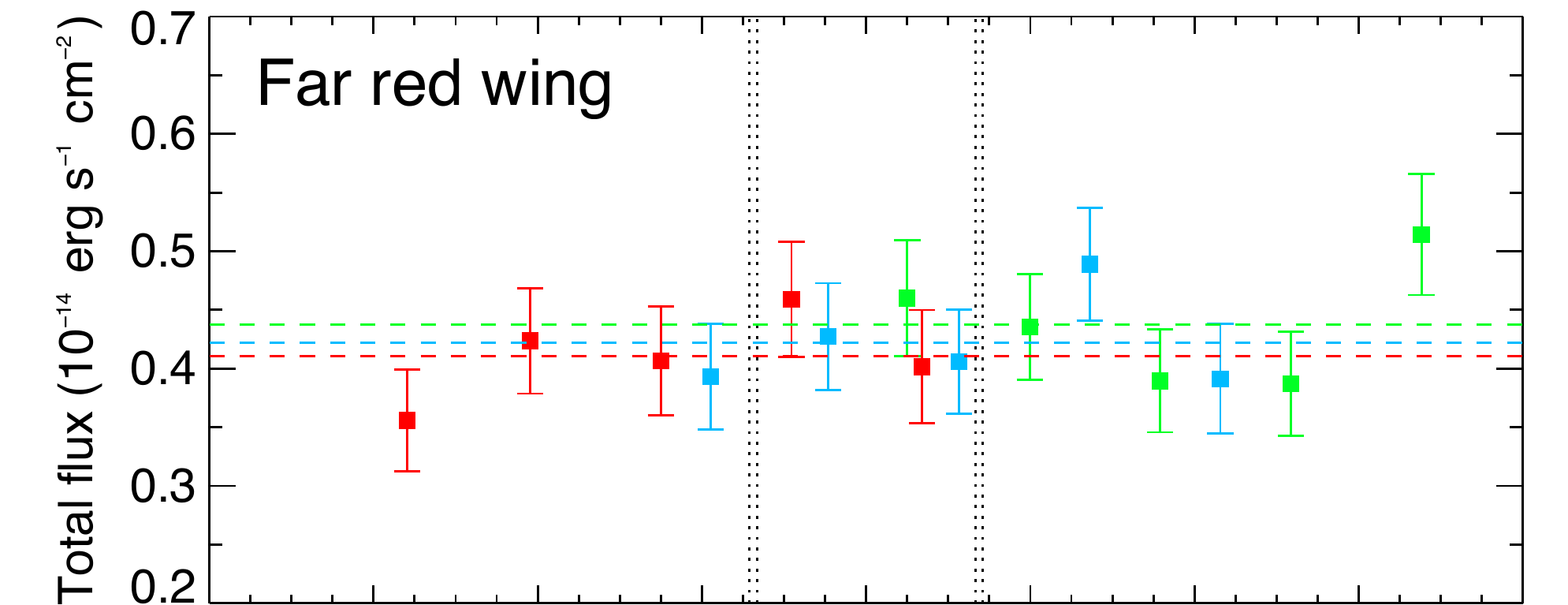}
\includegraphics[trim=2.5cm 2.9cm 7cm 10.3cm,clip=true,width=\columnwidth]{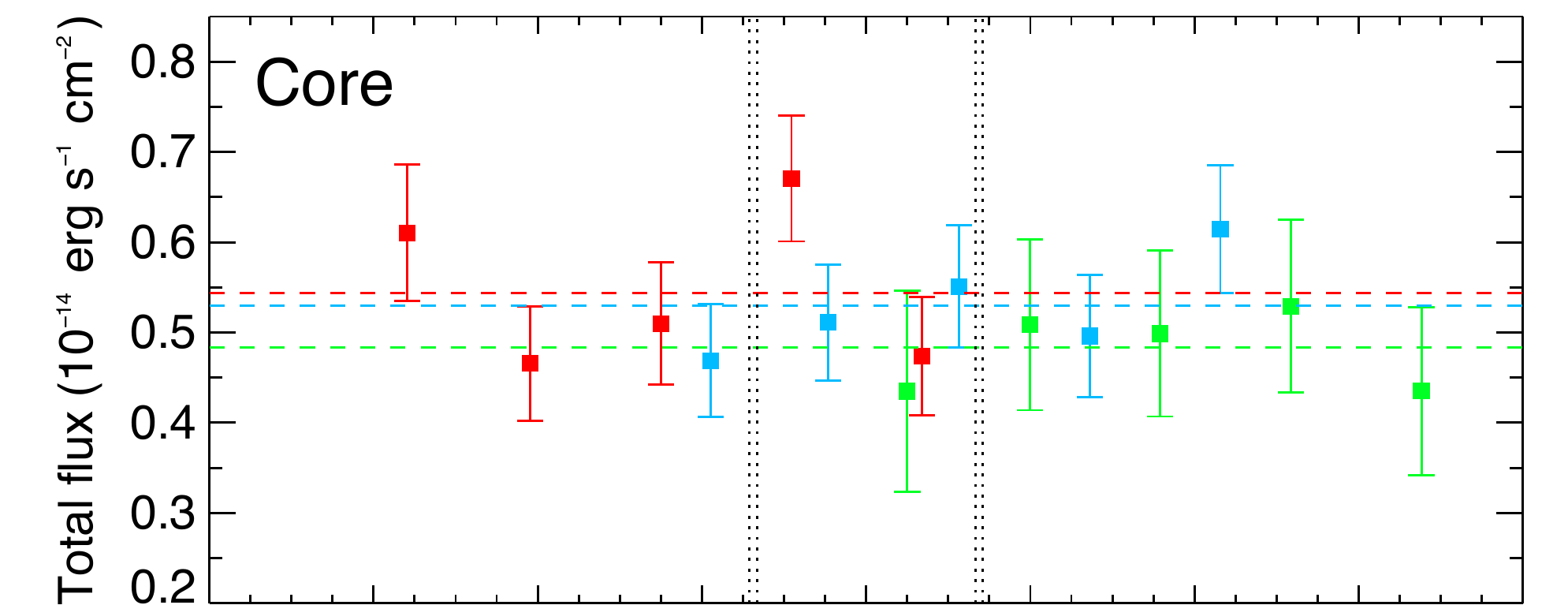}
\includegraphics[trim=2.5cm 2.9cm 7cm 10.3cm,clip=true,width=\columnwidth]{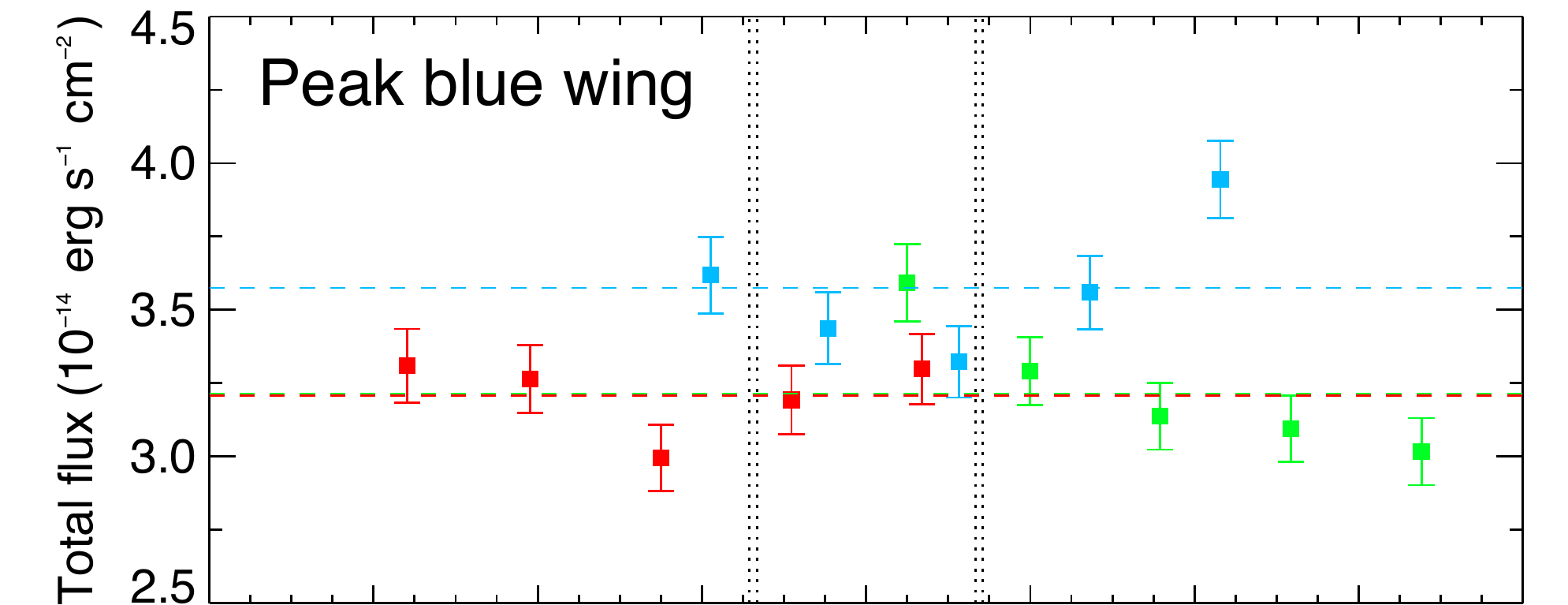}
\includegraphics[trim=2.5cm 1cm 7cm 10.3cm,clip=true,width=\columnwidth]{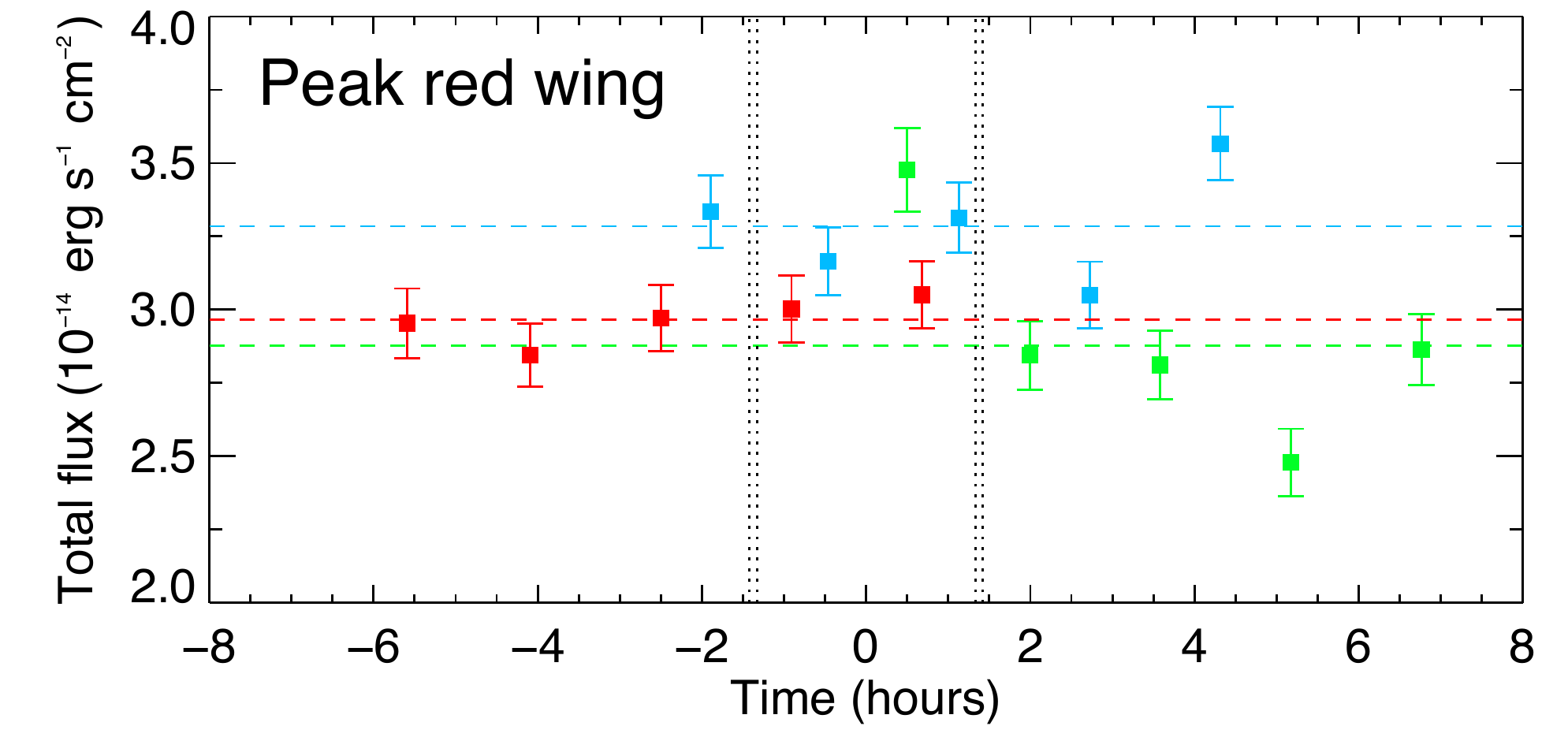}

\caption[]{Comparison of the Lyman-$\alpha$ flux between Visits 1 (blue), 2 (green), and 3 (red), integrated in complementary bands. All spectra have been interpolated over a common wavelength table before being summed, from top to bottom, in [-300 ; -183]\,km\,s$^{-1}$, [183 ; 300]\,km\,s$^{-1}$, [-33 ; 33]\,km\,s$^{-1}$, [-183 ; -39]\,km\,s$^{-1}$, and [39 ; 183]\,km\,s$^{-1}$. Squares correspond to the spectra measured during each HST orbit, and dashed lines to the reference spectra.}
\label{fig:light_curves}
\end{figure}

Our goal in this section is to characterize the variability between the exposures of each visit, and to search for transit signatures of an extended atmosphere around HD\,97658 b. This was done by calculating the ratio of each exposure spectrum to the reference spectrum of the corresponding visit. The S/N per pixel is too low in a 300\,s exposure to study spectral features, and we used the spectra averaged over the full duration of each HST orbit. We identified features characterized by relative flux variations with S/N larger than 2, and extending over more than three pixels (i.e., we excluded intervals narrower than STIS spectral resolution, about 0.1\,\AA). Using the bootstrap-like method detailed in \citet{Bourrier2013}, we also calculated the false-positive probability (p-value) to have features as significant as the ones we detected but caused by noise alone. The results of this analysis are given in Table~\ref{tab:signatures}.\\
Most known evaporating giants are characterized by an extended upper atmosphere of neutral hydrogen partially filling or overflowing the planetary Roche lobe. Gravity and interactions with the host star (in particular radiation pressure and charge-exchange) shape the escaping gas from a dense, low-velocity coma around the planet into a diluted, fast-moving cometary tail trailing behind the planet. Observationally, this atmospheric structure results in a deep absorption in the core of the Lyman-$\alpha$ line caused by the transit of the coma, and the absorption decreases and shifts to negative velocities in the blue wing of the Lyman-$\alpha$ line when the exosphere later transits the star. We found none of these features in our analysis of HD\,97658 Lyman-$\alpha$ line transit observations. While deviations with respect to the reference spectrum were detected in each visit, most of them have S/N lower than 3 and p-values up to 70\% (Table~\ref{tab:signatures}). Even if an exosphere-absorbed spectrum had been included in the weighted mean of the spectra used to build the reference spectrum, residual variations should still occur in the spectral range of this absorption signature. Instead, the measured variations do not show any correlation from one spectrum to another: they arise in all parts of the line, at no specific time relative to the planet transit, and with no consistency between the different epochs. \\
Interestingly, as can be seen in Table~\ref{tab:signatures}, Visit 1 is more variable than the other visits and  Visit 2 displays a significant variation (S/N$>$3\,$\sigma$, p-value$<$5\%) in the red wing. To further investigate this short-term variability we compare in Fig.~\ref{fig:light_curves} the light curves obtained by integrating the flux in the five bands identified in Sect.~\ref{sect:longterm_var}: the blue and red ``far'' wings, the core of the observed line, and its peaks. Within the statistical variations, the flux is stable in the far wings and the core of the line, not only between the different epochs, but also over a timescale of a few hours. In contrast, the regions of higher flux at the peaks of the observed line display significant systematic variations. Given that these spectral bands are the same that show long-term variability between Visit 1 and the other epochs (Sect.~\ref{sect:longterm_var}), this short-term variability could also be caused by activity in the hot upper chromosphere, stronger in Visits 1 and 2.\\
We thus conclude that there are no detectable transit signatures of an extended atmosphere around HD\,97658 b. The changes observed between the different spectra in each visit arise mainly from statistical variations around the reference spectrum, possibly from intrinsic variability of the stellar chromosphere, and in any case with no variations consistent with a transit light curve.\\

\begin{table}
\caption{Flux deviations from HD\,97658 reference Lyman-$\alpha$ spectra. In each visit, the upper subpanel corresponds to the blue wing of the line, the lower subpanel to the red wing.}
\centering
\begin{tabular}{cccccc}
\hline
\hline
\noalign{\smallskip}    
Visit    & Orbit & Spectral range & Amplitude   & S/N  & p-value        \\      
\noalign{\smallskip}
\hline
\hline
\noalign{\smallskip}
1       &   0   & \multicolumn{3}{c}{N/D}  \\ 
        &   1   & [-264 ; -238] &  -33.6$\pm$17.6\%   & 2.0   & 71\%  \\ 
        &   2   & [-251 ; -67 ] &  -10.5$\pm$3.7\%    & 2.9   & 12\%  \\ 
        &   3   & \multicolumn{3}{c}{N/D}  \\  
        &   4   & [-185 ; -27] &  10.3$\pm$4.0\%   & 2.6   & 26\%  \\ 
        &       & [-264 ; -211] &  48.6$\pm$22.2\%   & 2.2   & 49\%  \\ 
\hline
        &   0   & [157 ; 236] & -18.9$\pm$9.4\% & 2.0 & 63\%   \\ 
        &   1   & [78 ; 118] & -12.0$\pm$5.8\% & 2.1 & 59\%   \\ 
        &   2   & \multicolumn{3}{c}{N/D}  \\ 
        &   3   & [131 ; 184] & -20.0$\pm$7.8\% & 2.6 & 25\%  \\ 
        &   4   & [65 ; 118] & 13.4$\pm$5.8\%  & 2.3  &  39\%  \\               
\hline
\hline                  
\noalign{\smallskip} 
2       &   0   & [-184 ; -39]   &  11.9$\pm$4.5\%   & 2.6  & 20\%  \\ 
        &   1   & \multicolumn{3}{c}{N/D}  \\ 
        &   2   & \multicolumn{3}{c}{N/D}  \\ 
        &   3   & [-197 ; -158]  &  28.6$\pm$10.4\%   & 2.8  & 15\% \\  
        &   4   & [-66 ; -26.]  &  14.6$\pm$7.0\%   &  2.1  &  55\% \\  
\hline
        &   0   & [40 ; 184]   &  20.9$\pm$5.5\%   & 3.8  &  0.6\%  \\ 
        &   1   & \multicolumn{3}{c}{N/D}  \\ 
        &   2   & \multicolumn{3}{c}{N/D}  \\ 
        &   3   & [40 ; 171]   &  -14.4$\pm$4.4\%   & 3.3 &  3.7\%  \\ 
        &   4   & \multicolumn{3}{c}{N/D}  \\   
\hline
\hline                  
\noalign{\smallskip}
3       &   0   & \multicolumn{3}{c}{N/D}  \\ 
        &   1   & \multicolumn{3}{c}{N/D}  \\ 
        &   2   & [-113 ; -34]      &  -11.4$\pm$4.5\%   & 2.5 &  26\%  \\
        &       & [-245 ; -179]      &  -24.7$\pm$11.1\%   & 2.2  &  47\%  \\      
        &   3   & \multicolumn{3}{c}{N/D}  \\ 
        &   4   & \multicolumn{3}{c}{N/D}  \\ 
\hline
        &   0   & [150 ; 282]      &  -18.3$\pm$8.5\%   & 2.2 &  52\%  \\ 
        &   1   & \multicolumn{3}{c}{N/D}  \\ 
        &   2   & \multicolumn{3}{c}{N/D}  \\  
        &   3   & [203 ; 242]      &  49.1$\pm$24.5\%   & 2.0 &  64\%  \\ 
        &   4   & \multicolumn{3}{c}{N/D}  \\ 
\noalign{\smallskip}
\hline
\hline
\multicolumn{6}{c}{Note: N/D indicates that no variations were found above 2$\sigma$}\\
\multicolumn{6}{c}{from the reference spectrum of the visit. Spectral ranges are}\\
\multicolumn{6}{c}{given as velocities in the star rest frame (km\,s$^{-1}$), and flux}\\ 
\multicolumn{6}{c}{variations are in erg\,s$^{-1}$\,cm$^{-2}$\,\AA$^{-1}$.} \\
\end{tabular}
\label{tab:signatures}
\end{table}

\section{Intrinsic Lyman-$\alpha$ line}
\label{sec:recons}

The observed Lyman-$\alpha$ line is affected by absorption from the ISM, which needs to be corrected for in order to estimate the intrinsic stellar Lyman-$\alpha$ line profile. Knowledge of the intrinsic line is useful for several reasons: it is proportional to the stellar radiation pressure, which is essential in understanding the structure of a planetary exosphere (Sect.~\ref{sec:intrinsic_values}); intrinsic lines in different epochs can be directly compared, even when they are derived from observations obtained with different instrumental settings (Sect.~\ref{sec:intrinsic_values}); and empirical relations between the intrinsic Lyman-$\alpha$ flux and the stellar EUV spectrum can be used to estimate the irradiation of the planet (Sect.~\ref{sec:XUV_spec}). \\

\subsection{Reconstruction method}

The intrinsic Lyman-$\alpha$ line can be reconstructed directly from the STIS spectra, using the method detailed in \citet{Bourrier2015_GJ436} (see, e.g., \citealt{Wood2005}; \citealt{Ehrenreich2011}; \citealt{Youngblood2016}). A model profile of the stellar Lyman-$\alpha$ line is absorbed by the interstellar hydrogen and deuterium, and convolved by the STIS line spread function (LSF). A Metropolis-Hasting Markov chain Monte Carlo algorithm is used to find the best fit between the model spectrum and the observed spectrum, which are compared over the entire spectral line (see the spectral ranges in Table~\ref{tab:results_la}). The medians of the posterior probability distributions are chosen as the final best-fit values for the model parameters, and their 1$\sigma$ uncertainties are obtained by finding the intervals on both sides of the median that contain 34.15\% of the accepted steps.\\
In addition to the reference spectra calculated in Sect.~\ref{sec:line_variations} for Visits 1, 2, and 3, we used the measurement of HD\,97658 Lyman-$\alpha$ line secured by \citet{France2016} in February 2015, about a quarter orbital period after the transit (hereafter, Visit 0). Because it was obtained with STIS E140M mode this spectrum could not be directly compared with our G140M measurements in the previous sections. Nonetheless, it can be used to reconstruct the intrinsic line at another epoch, and the higher resolution of the echelle mode combined with the use of the 0.2''$\times$0.06'' slit significantly limits the airglow contamination and enables the measurement of the ISM thermal broadening, constrained by the ISM deuterium absorption (Fig.~\ref{fig:intrinsic_line}). \\
The Lyman-$\alpha$ line can be double-peaked (e.g., the K dwarf HD\,189733; \citealt{Bourrier2013}) or single-peaked (e.g., the M dwarf GJ\,436;  \citealt{Ehrenreich2015}), depending on the presence of a self-reversal arising from the absorption by the upper chromospheric layers of the Lyman-$\alpha$ emission from the lower chromosphere. We thus tried different combinations of Gaussian and Voigt profiles to model the possible presence of a self-reversal (\citealt{Bourrier2015_GJ436}; \citealt{Youngblood2016}). The fit of the STIS LSF is included in the reconstruction process because there is no published tabulated LSF profile for the 50''$\times$0.05'' slit used in Visits 1 to 3, and because it can be contaminated by systematic noise sources (e.g., stray light). Past studies (e.g., \citealt{Ehrenreich2012}; \citealt{Bourrier2013}; \citealt{Bourrier2015}) have shown that the fitted LSF can be wider than the 50''$\times$0.1'' slit LSF, and can possess broad, shallow wings. We thus modeled the LSF as a Gaussian core, and tried including an additional Gaussian representing these wings. We used the Bayesian Information Criterion (BIC; \citealt{deWit2012}) to prevent over-fitting in our search for the best model for the Lyman-$\alpha$ and LSF profiles. \\

\subsection{ISM and line properties}
\label{sec:intrinsic_values}

In a first step we performed the reconstruction independently for each visit, and found that the lowest $\chi^2$ and BIC was always obtained when the intrinsic line is modeled as a single-peaked Voigt profile. Surprisingly, \citet{Bourrier2013} found that the intrinsic line of HD\,189733, which has the same stellar type as HD\,97658, was best reconstructed by including a self-reversal. Furthermore, the Mg\,II line of HD\,97658 (\citealt{France2016}) does present a self-reversed emission. The absence of a self-absorption at the peak of HD\,97658 Lyman-$\alpha$ line would be related to the temperature and density structure of neutral hydrogen in the upper chromosphere of the K dwarf (e.g., \citealt{Curdt2001}).\\

In Visit 0, a single Gaussian best fits the echelle mode LSF, while in the three visits obtained with the G140M the LSF is best modeled with a double-Gaussian profile. The free parameters of the fit are thus, for each epoch, the heliocentric velocity of the star $\gamma_{*}$, the Lyman-$\alpha$ line profile parameters (maximum stellar flux at 1\,au $f_\mathrm{peak}$(1\,au), Doppler width $\Delta$\,v$_{D}$ and damping parameter $a_\mathrm{damp}$), and the LSF profile parameters (the width of the core $\sigma^\mathrm{core}_\mathrm{LSF}$, and when required the width of the wings $\sigma^\mathrm{wing}_\mathrm{LSF}$ and ratio between the peaks of the core and the wings $r^{\mathrm{core}/\mathrm{wing}}_\mathrm{LSF}$). The ISM parameters for neutral hydrogen are the column density log$_{10}$\,$N$(H\,{\sc i}), Doppler broadening parameter $b$(H\,{\sc i}), and velocity relative to the star $\gamma$(H\,{\sc i})$_{/*}$. The LISM Kinematic Calculator\footnote{\mbox{\url{http://sredfield.web.wesleyan.edu/}}} (\citealt{Redfield_Linsky2008}) predicts that the line of sight (LOS) toward HD\,97658 passes within 20$^{\circ}$ of the LIC, Leo, and NGP clouds. We tried including two ISM components in the reconstruction process, but found that the second component could not be well constrained. We thus modeled the ISM opacity from a single component as the combination of two Voigt profiles for the atomic hydrogen and deuterium, separated by about 0.33\,\AA. We used a D\,{\sc i}/H\,{\sc i} ratio of 1.5$\times$10$^{-5}$\ ({e.g.}, \citealt{Hebrard_Moos2003}, \citealt{Linsky2006}) and assumed there is no turbulent broadening (implying that $b$(D\,{\sc i}) = $b$(H\,{\sc i})/$\sqrt{2}$, \citealt{Wood2005}). Assuming that the ISM kept the same properties over the year of observations of HD\,97658, we used the weighted mean of the spectra of  Visits 1, 2, and 3  as a reference to estimate the ISM column density and radial velocity (Visit 0 was not included because of its different instrumental response). Shifted to the heliocentric rest frame, the derived ISM velocity (7.2$\pm$0.3\,km\,s$^{-1}$) is in between the predicted velocities for the Leo and NGP clouds  (4.7$\pm$0.8\,km\,s$^{-1}$ and 9.5$\pm$0.8\,km\,s$^{-1}$). Based on the Visit 0 spectrum alone, \citet{Youngblood2016} found a degeneracy between a low $b$(H\,{\sc i}) value (6\,km\,s$^{-1}$), high column density (18.45) solution and a high $b$(H\,{\sc i}) value (12\,km\,s$^{-1}$), low column density (18.2) solution. We thus confirm their expectation that the second solution is more realistic, and that the ISM component along HD\,97658 LOS has a high Doppler broadening (13.1$\stackrel{+1.3}{_{-1.8}}$\,km\,s$^{-1}$) and a low column density (18.18$\stackrel{+0.04}{_{-0.03}}$). The parameters log$_{10}$\,$N$(H\,{\sc i}) and $\gamma$(H\,{\sc i})$_{/*}$ were set to these best-fit values in the final reconstructions. ISM Doppler broadening $b$(H\,{\sc i}) cannot be constrained by the low-resolution G140M spectra and was estimated through the reconstruction of the Visit 0 echelle spectrum, along with the LSF and intrinsic line properties at this epoch. We set $b$(H\,{\sc i}) to its derived best-fit value and reconstructed the intrinsic line for Visits 1, 2, and 3 independently.\\

\begin{figure*}
\centering
\begin{minipage}[b]{\textwidth}   
\includegraphics[trim=4.2cm 5.7cm 3.8cm 1.9cm,clip=true,width=0.511\columnwidth]{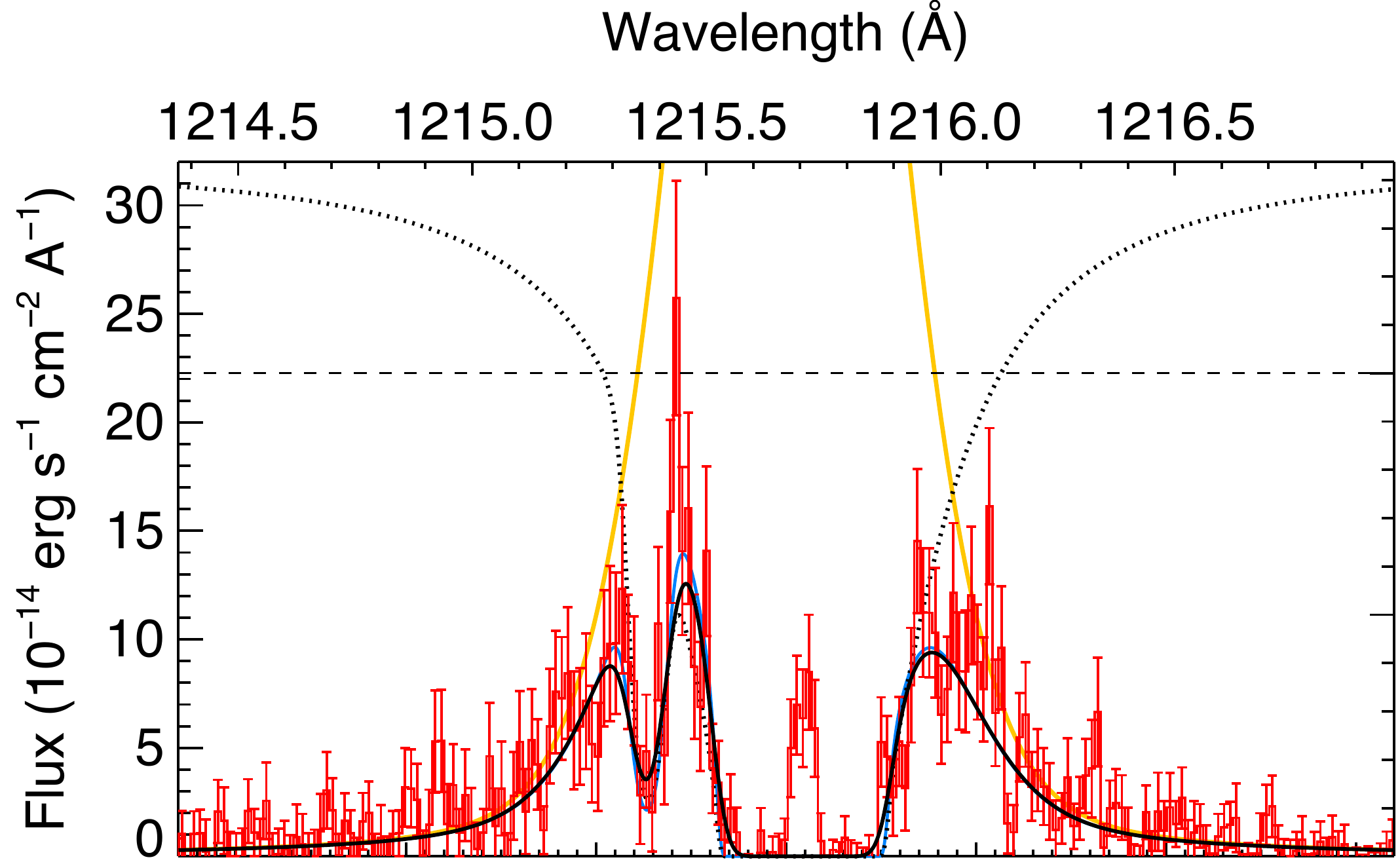}
\includegraphics[trim=6.71cm 5.7cm 2.2cm 1.9cm,clip=true,width=0.49\columnwidth]{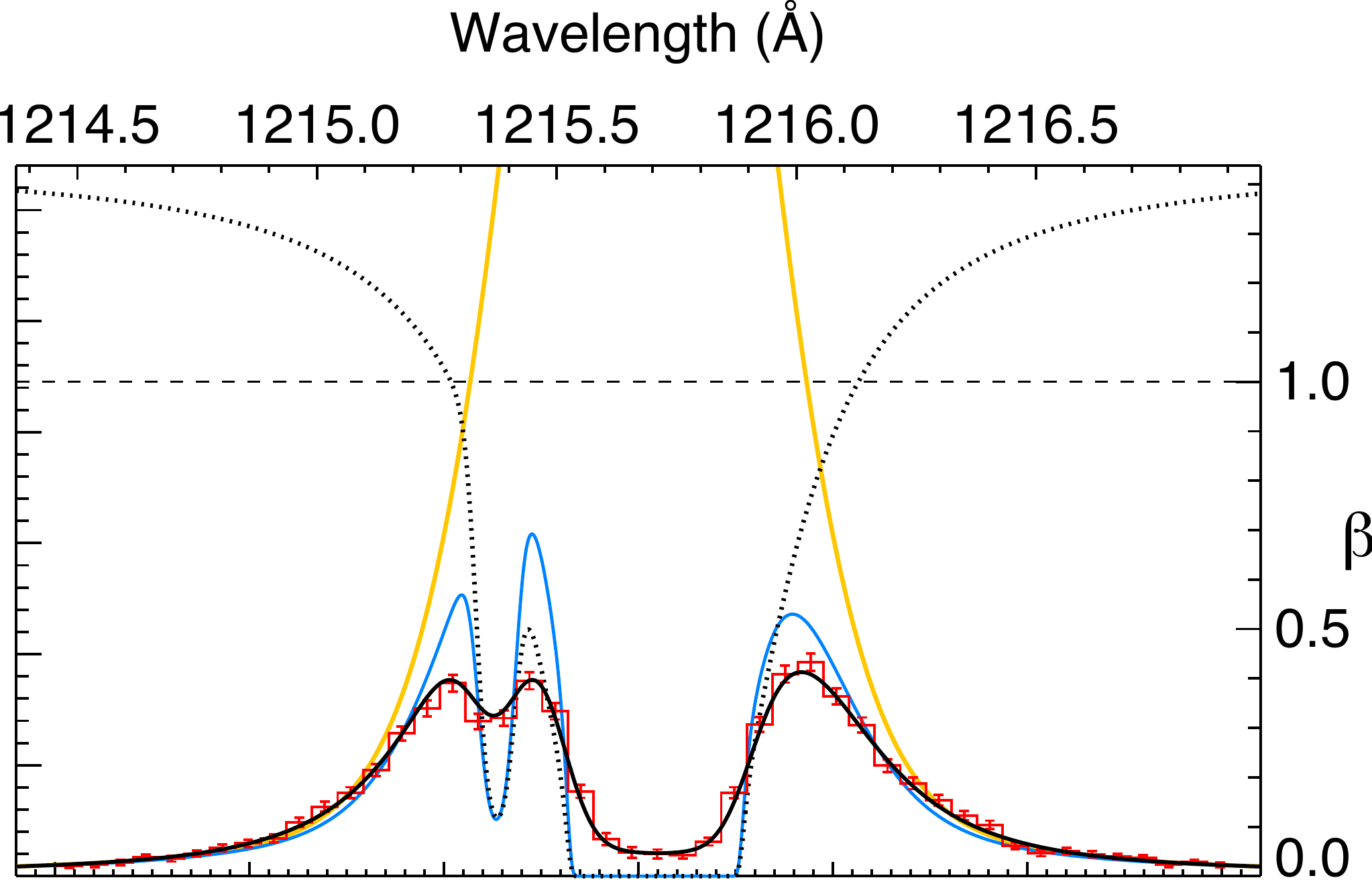}\\
\includegraphics[trim=4.2cm 4.7cm 3.8cm 4.2cm,clip=true,width=0.511\columnwidth]{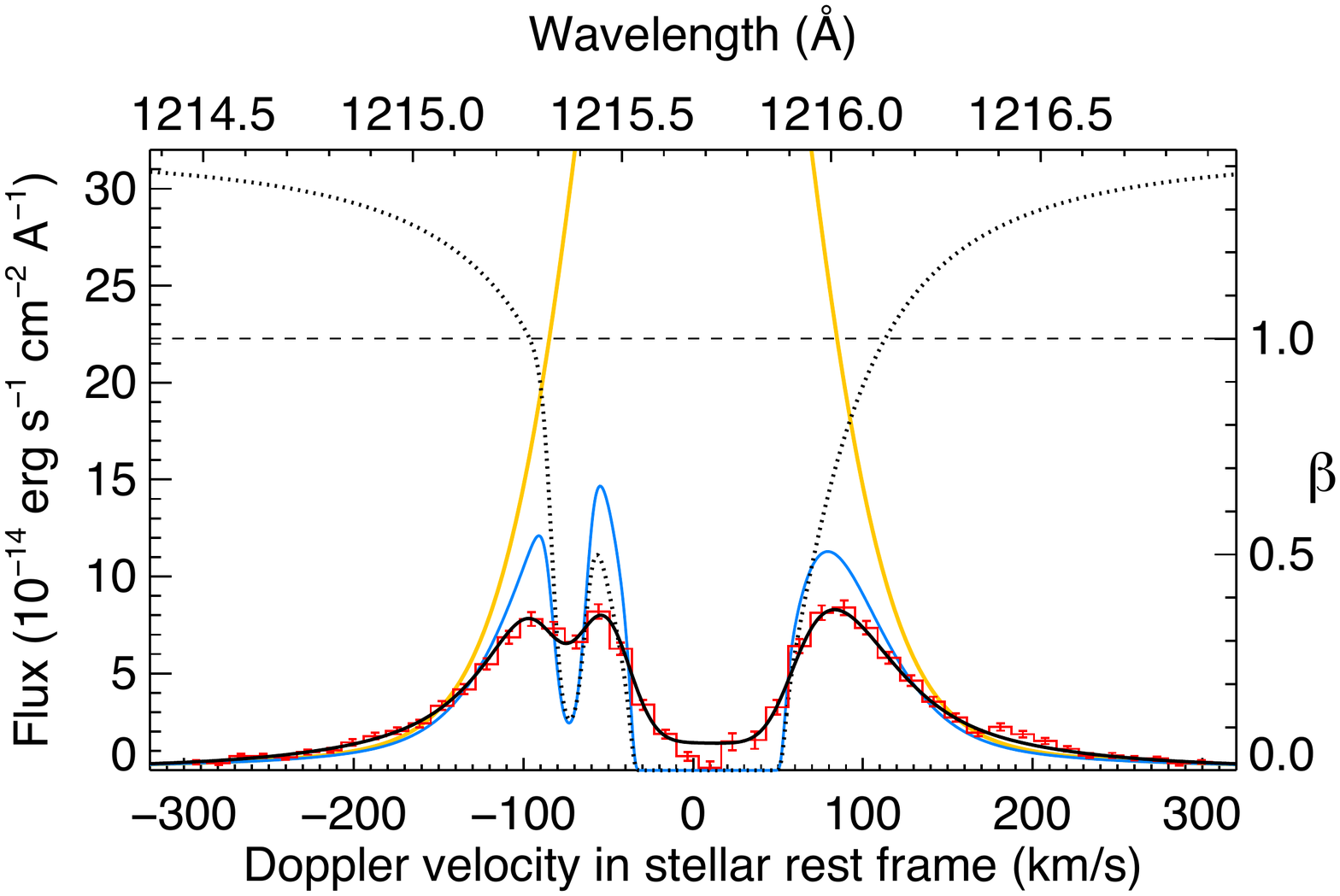}
\includegraphics[trim=6.71cm 4.7cm 2.2cm 4.2cm,clip=true,width=0.49\columnwidth]{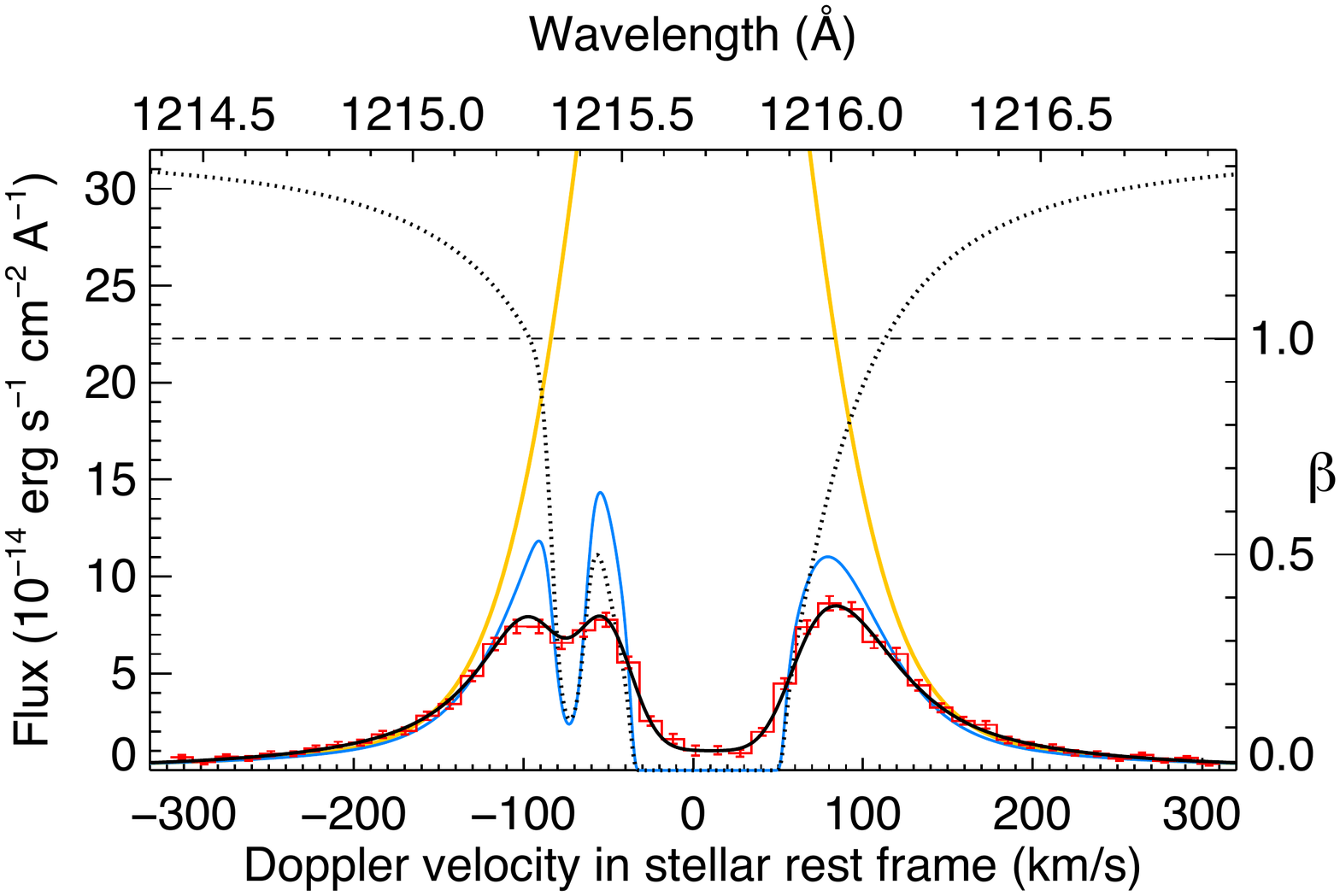}\\
\includegraphics[trim=-6cm 3.7cm 3.8cm 4.2cm,clip=true,width=0.739\columnwidth]{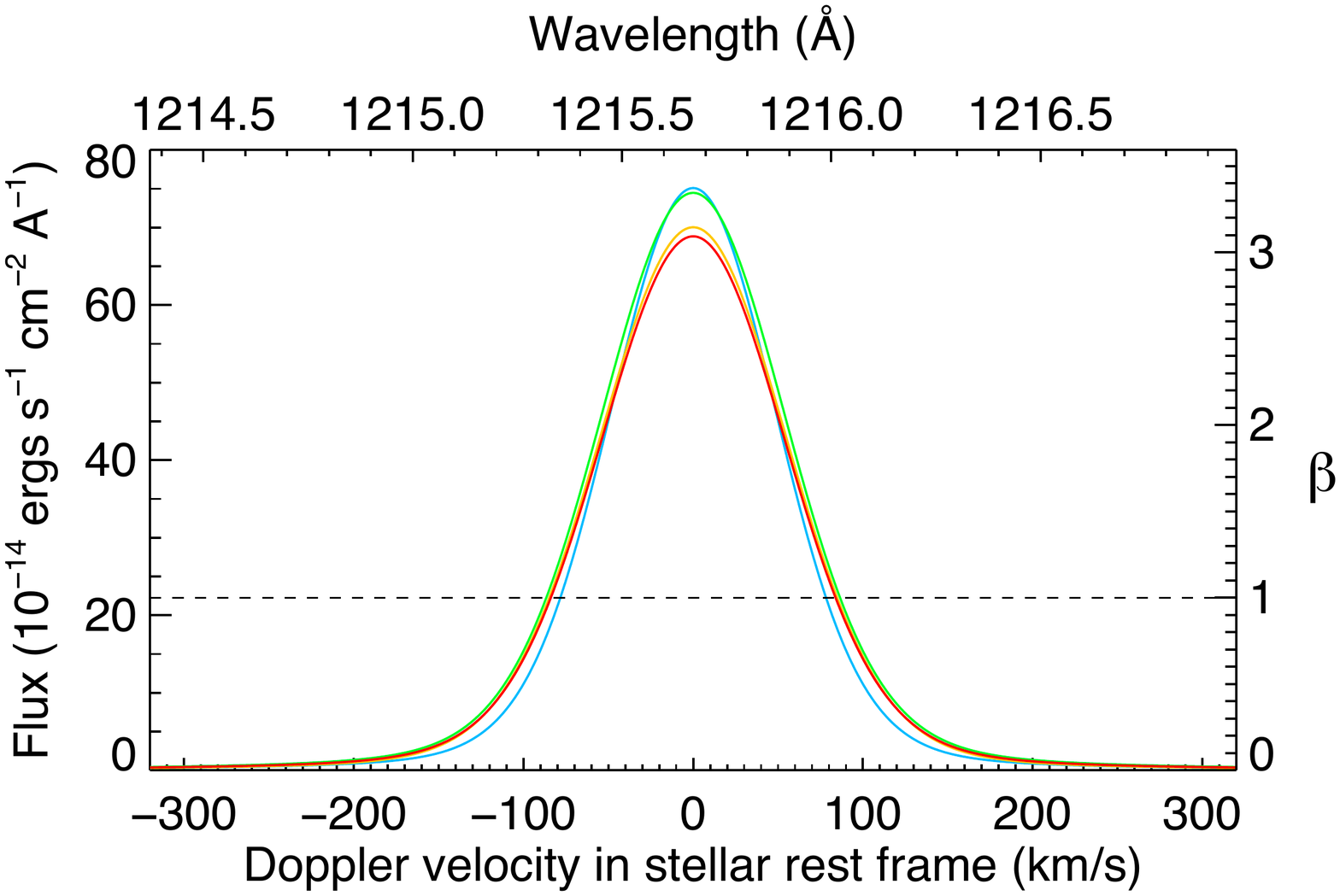}
\includegraphics[trim=26cm 3.7cm 2.2cm 4.2cm,clip=true,width=0.0348\columnwidth]{fit_theo_spectrum_all}\\
\end{minipage}
\caption[]{Lyman-$\alpha$ line profiles of HD\,97658 for Visits 0 (top left), 1 (top right), 2 (middle left), and 3 (middle right). The yellow line shows the theoretical intrinsic stellar emission line profile as seen by the planet upper atmosphere, scaled to the Earth distance. It also corresponds to the profile of the ratio $\beta$ between radiation pressure and stellar gravity (values reported on the right axis). The solid blue line shows the Lyman-$\alpha$ line profile after absorption by the interstellar hydrogen (1215.6\AA) and deuterium (1215.25\AA), whose cumulated profile is plotted as a dotted black line (ISM absorption in the range 0-1 has been scaled to the vertical axis range). The solid black line shows the line profile convolved with the high-resolution E140M LSF in Visit 0 and with the lower resolution G140M LSF in the other visits. It is compared to the observations shown as a red histogram. Note the residual airglow in the core of Visit 0 spectrum. Intrinsic profiles reconstructed for all visits can be compared in the bottom panel (Visit 0 in blue; Visit 1 in green; Visit 2 in yellow; Visit 3 in red).} 
\label{fig:intrinsic_line}
\end{figure*}

The best-fit reconstructed Lyman-$\alpha$ stellar profiles are displayed in Fig.~\ref{fig:intrinsic_line} and the corresponding model parameters are given in Table~\ref{tab:results_la}. The best-fit LSF is stable in the three G140M visits. The quality of the fits is good in all epochs, with reduced $\chi^2$ lower than 1.1. For Visits 2 and 3 the reduced $\chi^2$ were found to be lower than 1, and the uncertainties on the parameters derived from their fits can be considered as conservative. These results confirm that the mean spectra used as reference (Sect.~\ref{sec:line_variations}) trace  the intrinsic stellar line well. We do not detect significant variations between the four different epochs\footnote{The heliocentric velocity of the star shows a high dispersion between the visits, but this is likely caused by variations in the wavelength calibration of the spectra.}. Nonetheless, it is noteworthy that the median value for the peak amplitude of the line is higher in Visits 0 and 1, separated by only 2 months, and that the line-integrated flux is about 2\,$\sigma$ higher in Visit 1 than in later visits distant by more than 8 months (Table~\ref{tab:results_la}; Fig.~\ref{fig:intrinsic_line}). This is consistent with the higher flux measured at the peaks of the observed line in Visit 1 (Sect.~\ref{sect:longterm_var}). Its origin is likely an increased emission in the line core from a more active upper stellar chromosphere. \\
Finally, we note that the intrinsic Lyman-$\alpha$ line of HD\,97658 is strong enough that hydrogen atoms escaping from the planet would be accelerated away from the star by radiation pressure. Similarly to HD\,189733, this force can indeed overcome stellar gravity up to a factor of $\sim$3 (Fig.~\ref{fig:intrinsic_line}), although the different shapes of the intrinsic lines and the larger orbital distance of HD\,97658b suggests that the dynamics of a putative exosphere would be different from that of HD\,189733b. \\
                                                        
\begin{table*}
\caption{Parameters derived from the reconstruction of the Lyman-$\alpha$ line profiles.}
\begin{tabular}{lccccl}
\hline
\hline
\noalign{\smallskip} 
\textbf{Parameter} & \multicolumn{4}{c}{\textbf{ISM properties}}   & \textbf{Unit}\\
\hline
log$_{10}$\,$N$(H\,{\sc i}) &  \multicolumn{4}{c}{18.18$\stackrel{+0.04}{_{-0.03}}$}  & cm$^{-2}$   \\
$b$(H\,{\sc i})    &  \multicolumn{4}{c}{13.1$\stackrel{+1.3}{_{-1.8}}$}    & km\,s$^{-1}$\\
$\gamma$(H\,{\sc i})$_{/*}$  & \multicolumn{4}{c}{8.6$\stackrel{+0.4}{_{-0.3}}$}   & km\,s$^{-1}$   \\
\noalign{\smallskip}
\hline
\hline
\noalign{\smallskip} 
\textbf{Parameter} & \textbf{Visit 0} & \textbf{Visit 1} & \textbf{Visit 2} & \textbf{Visit 3} & \textbf{Unit} \\                        
\noalign{\smallskip}
\hline
\hline
\noalign{\smallskip}
\textit{Fit range} & [1214.66 - 1215.57] & [1214.43 - 1216.89] & [1214.44 - 1215.61] & [1214.46 - 1216.91] & \AA\ \\
                                   & [1215.86 - 1216.67]        &                                        & [1215.77 - 1216.36] &                                                            & \AA\ \\
                               &                                                           &             & [1216.52 - 1216.89] &                                                            & \AA\ \\   
\textit{$\chi^2$} & 146.0 & 39.3 & 18.4  & 28.6 & \\
\textit{d.o.f}    & 132 & 39 & 33 & 39 & \\
\hline
\noalign{\smallskip}
$r^{\mathrm{core}/\mathrm{wing}}_\mathrm{LSF}$ & -  & 0.91$\stackrel{+0.03}{_{-0.05}}$ & 0.93$\stackrel{+0.02}{_{-0.03}}$ & 0.95$\stackrel{+0.01}{_{-0.02}}$ &    \\
$\sigma^\mathrm{core}_\mathrm{LSF}$ & 1.69$\stackrel{+0.56}{_{-0.54}}$ & 0.92$\pm$0.08 & 0.94$\stackrel{+0.10}{_{-0.08}}$ & 1.03$\pm0.07$   & STIS pixels$^\dagger$ \\
$\sigma^\mathrm{wing}_\mathrm{LSF}$ & -  & 3.8$\stackrel{+1.0}{_{-0.7}}$ & 5.7$\pm$1.3 & 5.8$\stackrel{+1.6}{_{-1.3}}$ & STIS pixels$^\dagger$ \\
$f_\mathrm{peak}$(1\,au)     & 14.2$\stackrel{+4.6}{_{-3.0}}$ & 14.1$\pm0.6$ & 13.2$\stackrel{+0.7}{_{-0.6}}$ & 13.0$\stackrel{+0.6}{_{-0.5}}$ & erg\,cm$^{-2}$\,s$^{-1}$\,\AA$^{-1}$\\    
$\Delta$\,v$_{D}$     & 64.3$\stackrel{+5.9}{_{-5.1}}$ & 69.89$\stackrel{+2.1}{_{-1.7}}$ & 72.7$\pm$3.4 & 71.0$\pm$2.5 &  km\,s$^{-1}$   \\                                          
$a_\mathrm{damp}$                        & 0.037$\stackrel{+0.008}{_{-0.007}}$ & 0.048$\pm$0.003 & 0.036$\pm$0.007 & 0.044$\pm$0.004 &  \\
$\gamma_{*}$         & -2.3$\stackrel{+1.0}{_{-1.1}}$ & -5.4$\pm$0.5 & -4.5$\pm$0.6 & -4.3$\pm$0.5 & km\,s$^{-1}$  \\
$F_{\mathrm{Ly}\alpha}$(1\,au) &  7.6$\stackrel{+1.7}{_{-1.1}}$ & 8.4$\pm$0.2 & 7.8$\pm$0.2 & 7.8$\pm$0.2 & erg\,cm$^{-2}$\,s$^{-1}$\\
$F_{\mathrm{Ly}\alpha}$(Earth)  & 4.0$\stackrel{+0.9}{_{0.6}}$ & 4.4$\pm$0.1 & 4.1$\pm$0.1 & 4.1$\pm$0.1 & $\times$10$^{-13}$\,erg\,cm$^{-2}$\,s$^{-1}$\\                               
\hline                                                                  
\hline
\multicolumn{6}{l}{Note: $F_{\mathrm{Ly}\alpha}$ is the flux integrated in the entire Lyman-$\alpha$ line, calculated at Earth distance and at 1\,au from the star.} \\
\multicolumn{6}{l}{Note: ISM properties are derived from the reconstruction of the weighted mean of the Visits 1, 2, and 3 spectra, and from} \\
\multicolumn{6}{l}{the single Visit 0 spectrum.} \\
\multicolumn{6}{l}{$\dagger$: A pixel width is about 0.015\AA\, for the E140M mode and 0.053\AA\, for the G140M.} \\
\end{tabular}
\label{tab:results_la}
\end{table*}

\section{XUV irradiation of the upper atmosphere}
\label{sect:X-EUV}

\subsection{Measurement of the X-ray stellar emission}
\label{sect:Xray_data}

\begin{figure}
\centering
\includegraphics[trim=0cm 0cm 0cm 0cm,clip=true,width=\columnwidth]{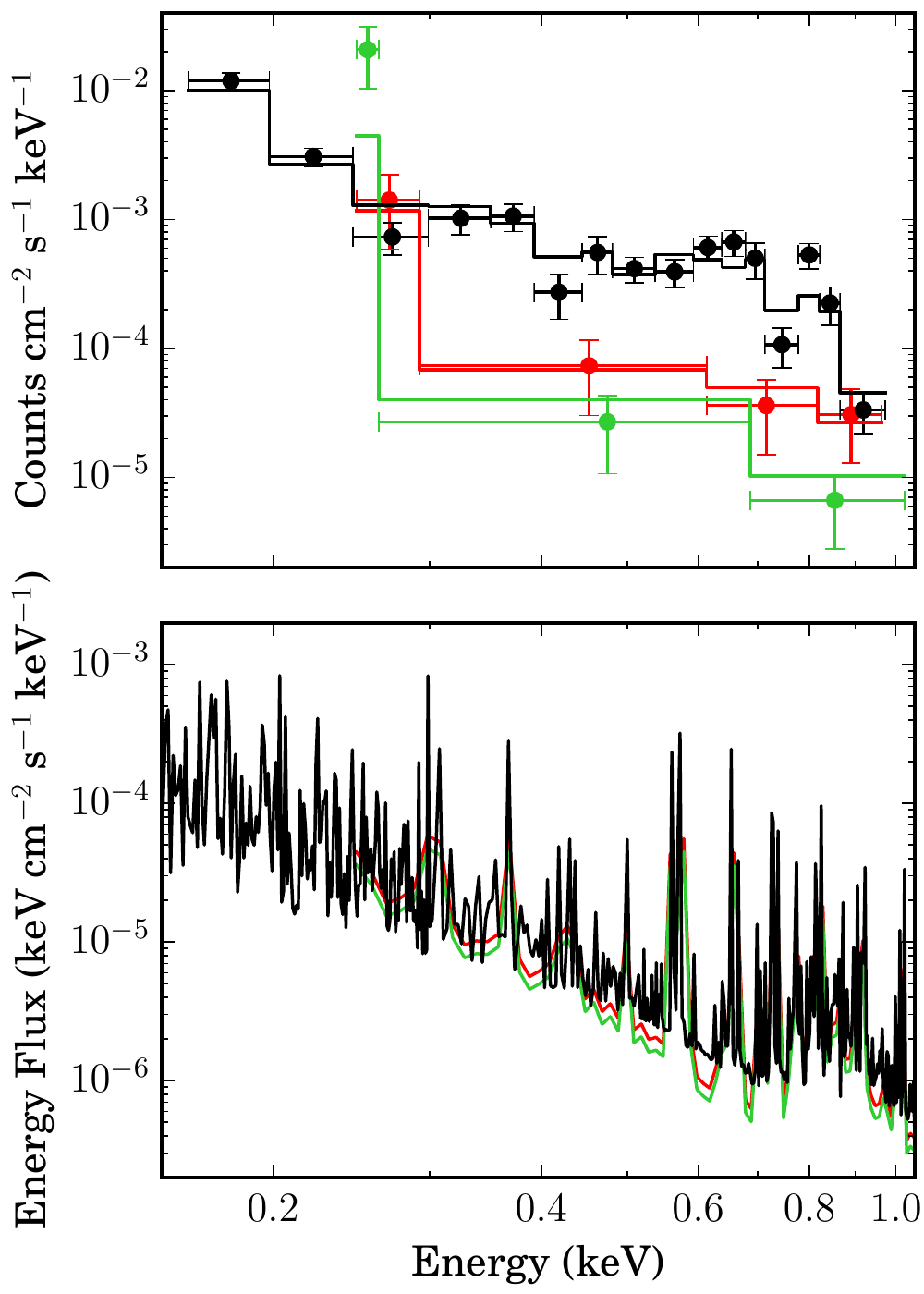}
\caption[]{X-ray spectrum of HD\,97658. The upper panel shows the spectra in units of measured counts, with XMM-Newton EPIC-PN in June 2015 (black points), and with Chandra in December 2015 (red points) and March 2016 (green points). To account for the different sensitivity of Chandra and EPIC-PN, the count rates are divided by the effective area of each instrument in the energy range covered by each bin. While this approach allows for a better comparison, it is an approximation that does not account for the redistribution matrix. The histogram spectra correspond to the best fits obtained with the CEMEKL model, which are plotted in flux units and displayed at high spectral resolution in the lower panel.}
\label{fig:Xray_spectra}
\end{figure}

We observed HD\,97658 (PI: P.J~Wheatley) in June 2015 with XMM-Newton/EPIC-PN (Obs ID: 0764100601; duration 33.9\,ks), and contemporaneously with the Lyman-$\alpha$ observations in December 2015 (Obs ID: 18724; duration 19.82\,ks) and March 2016 (Obs ID: 18725; duration 19.82\,ks) with Chandra/ACIS-S (Table~\ref{obs_log}). The XMM EPIC-PN spectrum shows HD\,97658 to be about ten times fainter and much softer than HD\,189733, even though both stars are about at the same distance ($\sim$20\,pc) and have the same stellar type. This indicates a significantly lower coronal activity for HD\,97658, possibly due to its older age (9.7$\pm$2.8\,Gyr, compared to 5.3$\pm$3.8\,Gyr for HD\,189733; \citealt{Bonfanti2016}). We use the flux measured in 0.15--2.4\,keV with XMM EPIC-PN, and 0.234--2.4\,keV with Chandra to fit the X-ray spectrum of the star at the different epochs of observations. We used C-statistics to fit spectral models to the data, a more reliable method than $\chi^2$ statistics for low count rates. The XMM EPIC-PN spectrum was thus binned to a minimum of 10 counts per bin, while the Chandra spectra were set to a minimum of 3 counts per bin. In a first step we tested the merits of three different spectral models by fitting only the XMM EPIC-PN spectrum. The first model (APEC; \citealt{smith2001}), generally used for X-ray observations of coronal emission from late-type stars, is a single-temperature model of optically thin thermal plasma. Here we use two single-temperature components in order to approximate a multi-temperature plasma. The second (CEMEKL; \citealt{Schmitt1990}; \citealt{Singh1996}) and third (C6PMEKL; \citealt{Lemen1989}; \citealt{Singh1996}) models both generate a spectrum from a continuous range of temperatures, with the emission measure distribution set respectively by a power law and a Chebyshev polynomial. The C6PMEKL model differs from the CEMEKL model by including the possibility for higher order terms. Fluxes derived from the three models showed good agreement in the observed energy band 0.2--2.4\,keV in which the PN camera is well calibrated. However, we need to extrapolate the X-ray spectrum down to 0.1\,keV (124\,\AA) to derive the stellar EUV flux with the relation from \citet{Chadney2015}. For our two-temperature APEC model fit the there is no constraint on the relative temperatures and emission measures of the two components, and this seems to allow too much freedom when extrapolating the fitted spectrum, with a very large flux implied between 0.1 and 0.2\,keV. We regard this extrapolation as unreliable. The emission measure distribution in the other two models is more tightly constrained and probably represents more closely the true emission measure distribution of the star, and yields more consistent and plausible results down to 0.1\,keV. Since the goodness of fit is marginally better for the simpler CEMEKL model, it was used  to fit the entire X-ray dataset.\\
We carried out joint fits of the XMM and Chandra spectra, forcing the spectral profile to be the same but allowing the flux level to vary between observations. We note that the Chandra ACIS instrument is accumulating a contaminant on the filter, which is reducing the soft-band sensitivity. Nonetheless, we used the most recent update of the Chandra calibration pipeline (CALDB 4.6.2; July 2014), which provides a realistic model for the contamination, and we checked that including soft energies in the simultaneous fit to the XMM and Chandra spectra did not change significantly the fit to the high energies alone. This gives us confidence in the derived Chandra fluxes, which were found to be consistent with each other but are up to twice as low as the XMM flux (at 2--2.5$\sigma$). Indeed, in the common Chandra band (0.243--2.4\,keV) we found fluxes at Earth distance of 2.18$\stackrel{+0.11}{_{-0.18}}\times$10$^{-14}$\,erg\,cm$^{-2}$\,s$^{-1}$ in June 2015 (XMM), 1.3$\pm$0.4$\times$10$^{-14}$\,erg\,cm$^{-2}$\,s$^{-1}$ in December 2015 (Chandra), and 1.1$\pm$0.4$\times$10$^{-14}$\,erg\,cm$^{-2}$\,s$^{-1}$ in March 2016 (Chandra). The stability of the Chandra measurements is consistent with that of the contemporaneous Lyman-$\alpha$ measurements in Visits 2 and 3. The spectra measured in each epoch are shown in Fig.~\ref{fig:Xray_spectra} along with their best fits that yield extrapolated fluxes between 0.1--2.4\,keV (5.17--124\,\AA\,) of 14.6$\stackrel{+2.4}{_{-2.5}}\times$10$^{-14}$\,erg\,cm$^{-2}$\,s$^{-1}$ in June 2015, 8.9$\stackrel{+2.5}{_{-2.8}}\times$10$^{-14}$\,erg\,cm$^{-2}$\,s$^{-1}$ in December 2015, and 7.2$\stackrel{+2.4}{_{-2.5}}\times$10$^{-14}$\,erg\,cm$^{-2}$\,s$^{-1}$ in March 2016.\\

\subsection{XUV irradiation}
\label{sec:XUV_irr}

\subsubsection{XUV spectrum}
\label{sec:XUV_spec}

We calculated the XEUV spectrum of HD\,97658, from 5\,\AA\, to 912\,\AA\,, to infer the photoionization rate of neutral hydrogen in the upper atmosphere of the planet and to constrain its energy-limited mass-loss rate. The EUV emission is mostly absorbed by the ISM and must be estimated through indirect methods.\\
In a first step, we estimated the flux in 124--912\,\AA\, for each epoch of observation using two independent methods. The EUV spectrum was estimated with the empirical relation from \citet{Chadney2015}, which is based on the high-energy X-ray flux measured in 5.17--124\,\AA\, (Sect.~\ref{sect:Xray_data}), and with the semi-empirical relations from \citet{Linsky2014},  which are based on the lower energy integrated intrinsic Lyman-$\alpha$ flux (Table.~\ref{tab:results_la}). We obtained the following fluxes at 1\,au from the star (in erg\,cm$^{-2}$\,s$^{-1}$): 4.26 (February 2015, from HST Visit 0), 4.87 (April 2015, from HST Visit 1), 6.57 (June 2015, from XMM), 4.44 and 4.93 (December 2015, from HST Visit 2 and Chandra, respectively), 4.37 and 4.36 (March 2016, from HST Visit 3 and Chandra, respectively). Estimating uncertainties on these EUV fluxes is not straightforward, as they depend on  uncertainties on the X-ray and the Lyman-$\alpha$ measurements, and on the empirical relations themselves. The median of the seven derived values yields a flux of 4.44\,erg\,cm$^{-2}$\,s$^{-1}$, and a rms dispersion of 19\%. Within this uncertainty we did not detect any trend in the EUV flux as a function of time, and although the stellar emission is higher in June 2015 it remains within 2.5\,rms of the median flux. The EUV emission of HD\,97658 can thus be considered  stable within 20\% variations, which shows not only that the \citet{Chadney2015} and \citet{Linsky2014} formulae can provide consistent first-order predictions of the EUV emission, but also that a flux of 4.44$\pm$0.84\,erg\,cm$^{-2}$\,s$^{-1}$ is a good estimate of HD\,97658 average emission in 124 - 912\,\AA\,. \\
Finally, we used the \citet{Linsky2014} relations to calculate the EUV spectrum in complementary wavelengths bands between 100 and 1170\,\AA\,, adjusting the global level of the fluxes so that they yield the value derived in 124--912\,\AA\,. For each band, we set an uncertainty of 19\%. The stellar emission in 5--100\,\AA\, was derived from the spectra obtained in Sect.~\ref{sect:Xray_data}, and fixed to the median value of the three epochs. Flux values for the entire XUV spectrum are given in Table.~\ref{tab:XEUV_flux}. \\

\begin{table}
\centering
\caption{XUV spectrum of HD\,97658}
\label{tab:XEUV_flux}
\begin{threeparttable}
\begin{tabular}{lcc}
\hline
\hline
\noalign{\smallskip}    
Wavelengths  & \multicolumn{2}{c}{Stellar flux}         \\      
\noalign{\smallskip}
\multicolumn{1}{c}{(\AA)}& \multicolumn{2}{c}{(erg\,s$^{-1}$\,cm$^{-2}$)}         \\      
\noalign{\smallskip}
 &  At 1\,au  &  At the planet (0.08\,au)       \\      
\noalign{\smallskip}
\hline
5 - 100         &   0.70      &  109.3                     \\ 
100 - 200       &   0.86   &   134.3                                       \\ 
200 - 300       &   0.89  &    139.6                       \\ 
300 - 400       &   1.03        &  160.7                           \\ 
400 - 500       &   0.08     &  12.4       \\  
500 - 600       &   0.29     &  44.7    \\ 
600 - 700       &   0.18      &   27.9                             \\ 
700 - 800       &   0.40      &   62.0                                     \\ 
800 - 912       &   0.92       &  144.3                            \\  
912 - 1170      &   0.74       &  115.6                                    \\ 
XUV (5 - 912) &   5.34        &  835.1            \\ 
\noalign{\smallskip}
\hline
\hline
\end{tabular}
  \begin{tablenotes}[para,flushleft]
  Note: The uncertainty is estimated to be 41\% on the X-ray flux and 19\% on the EUV fluxes.
  \end{tablenotes}
  \end{threeparttable}
\end{table}

\subsubsection{Photoionization rate}
\label{sec:photoion}

The photoionization rate per neutral hydrogen atom at the semi-major axis of the planet is defined as   
\begin{equation}
\Gamma_{\mathrm{ion}}=\int^{911.8\,\AA} \! \frac{F_\mathrm{XUV}(\lambda)\,\sigma_{ion}(\lambda)}{hc} \, \lambda\, \mathrm{d}\lambda, 
\end{equation} 
with $\Gamma_{\mathrm{ion}}$ in s$^{-1}$, $F_\mathrm{XUV}(\lambda)$ the stellar flux at 0.08\,au in erg\,s$^{-1}$\,cm$^{-2}$\,\AA$^{-1}$ and, $\sigma_{ion}$ the cross-section for photoionization in cm$^{2}$. The integration is performed up to the ionization threshold at 911.8\,\AA. The cross-section for hydrogen photoionization is wavelength-dependent, and we used the expression from \citet{Verner1996} and \citet{Bzowski2013},
\begin{equation}
\sigma_{\mathrm{ion}}=6.538\times10^{-32}\,\left( \frac{29.62}{\sqrt{\lambda}} +1 \right)^{-2.963}\,(\lambda-28846.9)^{2}\,\lambda^{2.0185},
\end{equation}
with $\sigma_{\mathrm{ion}}$ in cm$^2$ and wavelengths in \AA. Using the XUV spectrum given in Table~\ref{tab:XEUV_flux} and propagating its uncertainties, we obtained a photoionization rate at the semi-major axis of HD\,97658b of 4.8$\stackrel{+1.6}{_{-1.4}}\times$10$^{-5}$\,s$^{-1}$ (3.1$\stackrel{+1.0}{_{-0.9}}\times$10$^{-7}$\,s$^{-1}$ at 1\,au from the star). This corresponds to a photionization lifetime of about 5.8$\pm$1.8\,h at the semi-major axis.\\
\citet{Bourrier_lecav2013} found ionizing fluxes between 6 and 23 times the solar value in their study of HD\,189733 b. Using their conventions, this converts into photoionization rates at 1\,au in [4.6 - 17.7]$\times$10$^{-7}$\,s$^{-1}$, higher than the photoionization rates from HD\,97658 at the same distance and consistent with our aforementioned proposition that HD\,189733 has a more active upper chromosphere and corona. Since the hot Jupiter HD\,189733 b orbits about 2.5 times closer than HD\,97658b from its host star, the photoionization rate at HD\,189733 b is at least six times higher than at HD\,97658b. In contrast, \citet{Bourrier2016} found that the evaporating warm Neptune GJ\,436 b, in orbit around a colder M dwarf, is subjected to low photoionization rates in [1.4--3.4]$\times$10$^{-5}$\,s$^{-1}$ at the planet, explaining in part the formation of the giant exosphere of neutral hydrogen surrounding the planet (\citealt{Ehrenreich2015}). HD\,97658 b is thus closer to GJ\,436b in terms of photoionization, while being subjected to more than four times its radiation pressure (see \citealt{Bourrier2015_GJ436} and Fig.~\ref{fig:intrinsic_line} in the present paper). This strong radiation pressure would swiftly blow away neutral hydrogen atoms escaping HD\,97658b, which would then remain in this state for many hours owing to the low photoionization rates, leading eventually to the formation of a comet-like tail with a very long extension. We investigate in the next sections reasons that could explain our non-detection of such an exosphere.\\

\section{Observational constraints on the upper atmosphere}
\label{sec:EVE_sim}

\subsection{EVE simulations of HD\,97658 b}

We use the non-detection of neutral hydrogen absorption from HD\,97658 b (Sect.~\ref{sect:shortterm_var}) and the measurement of its radiation environment (Sect.~\ref{sec:XUV_irr}) to constrain the structure of its upper atmosphere. We interpret the transit observations of HD\,97658 b in the Lyman-$\alpha$ line  using the EVaporating Exoplanets (EVE) code (for a detailed description see, e.g., \citealt{Bourrier2015_GJ436}). The upper atmosphere is described in 3D with two different regions: the bottom layers are parameterized with a fixed set of atmospheric profiles, while Monte Carlo particle simulations are used to compute the dynamics of neutral hydrogen meta-particles in the exosphere. The transition between these two regions is set at $R_{\mathrm{trans}}$, which represents the altitude up to which the atmosphere keeps a global cohesion. For the atmosphere below the transition, hereafter called the isotropic atmosphere, we assume an upper atmosphere of pure neutral hydrogen with an isotropic, hydrostatic density profile. This profile is constrained by a mean temperature $T_{\mathrm{iso}}$ and scaled so that it matches the exospheric hydrogen density at $R_{\mathrm{trans}}$. To represent the expansion of the hydrogen outflow we set a radial velocity profile derived from Figure 3 in \citet{Koskinen2013a}, which increases from a few meters per second at the planet surface to a velocity $v_{\mathrm{trans}}$ at $R_{\mathrm{trans}}$. We note that the choice of this velocity profile does  not have much influence on our results compared to the value of $v_{\mathrm{trans}}$. At each time step, meta-particles representing groups of neutral hydrogen atoms with the same properties are released from the entire atmosphere at $R_{\mathrm{trans}}$ with a rate $\dot{M}_{\mathrm{H^{0}}}$. Because the orbit is slightly eccentric ($e$=0.078), the escape rate varies as the inverse square of the distance to the star, with a reference set at the semi-major axis. The initial velocity distribution of the particles relative to the planet is the combination of the outflow bulk velocity $v_{\mathrm{trans}}$, and an additional thermal speed component from a Maxwell-Boltzman velocity distribution defined by $T_{\mathrm{iso}}$. The dynamics of the particles is calculated in the stellar reference frame and results from the stellar and planetary gravities, the velocity-dependent stellar radiation pressure, and the inertial force linked to the non-Galilean stellar reference frame. Neutral hydrogen particles are also subjected to stellar photoionization, set by the rate 4.8$\times$10$^{-5}$\,s$^{-1}$ at the planet semi-major axis (as derived in Sect.~\ref{sec:photoion}) but varying with the particles' distance to the star. The impact of radiation pressure and photoionization onto neutral hydrogen particles is calculated taking into account self-shielding within the exospheric cloud.\\
The EVE code generates theoretical Lyman-$\alpha$ spectra as seen with HST/STIS after absorption by the planet, its upper atmosphere, and the interstellar medium using a limited number of parameters. The direct comparison of these spectra with observational data allows us to study the influence of $R_{\mathrm{trans}}$ (in $R_{\mathrm{pl}}$), $\dot{M}_{\mathrm{H^{0}}}$ (in g\,s$^{-1}$), $v_{\mathrm{trans}}$ (in km\,s$^{-1}$), and $T_{\mathrm{iso}}$ (in K). We create a single observational dataset by combining the spectra from Visits 2 and 3\footnote{To limit the effect of short-term variability, we used data averaged over the full duration of each HST orbit.}, which were shown to be very similar (Sect.~\ref{sect:longterm_var}). The exposures from these two visits do not overlap in terms of planetary orbital phase, except for the first orbit of Visit 3 and the last orbit of Visit 2, which were merged into a single exposure. These observations were compared to theoretical Lyman-$\alpha$ line spectra calculated every 5\,mn with EVE at a resolution $\Delta \lambda=$0.02\,\AA\, corresponding to $\Delta v=$5\,km\,s$^{-1}$ (about a quarter of the resolution of the STIS spectra at 1215.67\,\AA). The intrinsic line flux required to calculate the theoretical spectra and radiation pressure on hydrogen atoms was set to the average line between Visits 2 and 3 (Fig.~\ref{fig:intrinsic_line}). The merit function is the sum of the $\chi^2$ yielded by the comparison of the observed spectra with the theoretical spectra averaged during the time window of each observation, convolved with STIS LSF (set to the average reconstruction from Visits 2 and 3), and limited to the wavelength range corresponding to the velocity range [-300 ; 300]\,km\,s$^{-1}$.\\


\subsection{Structure of the upper atmosphere}

\begin{figure*}
\centering
\begin{minipage}[b]{\textwidth}   
\includegraphics[trim=1cm 0.5cm 0cm 0cm,clip=true,width=\columnwidth]{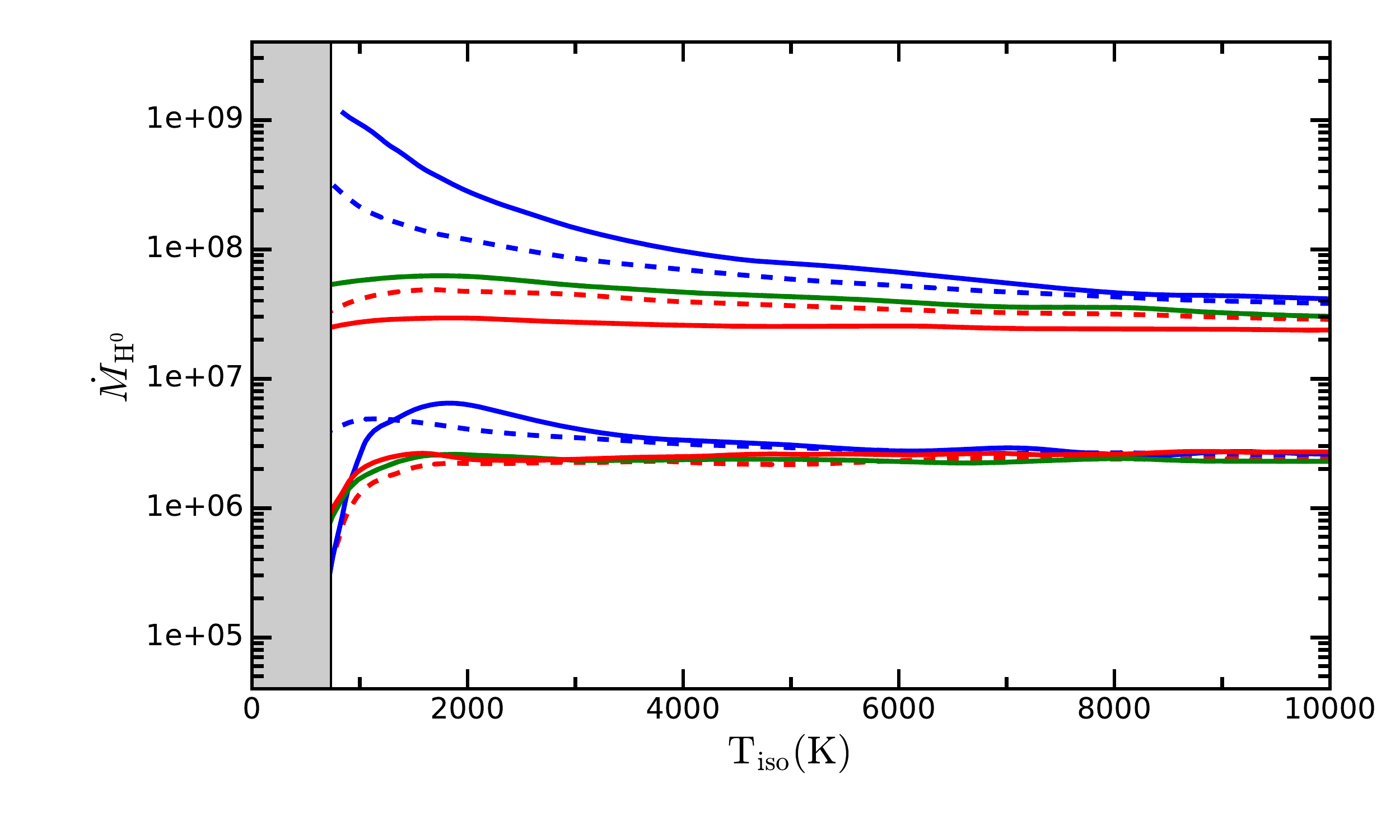}
\caption[]{Upper limit on the neutral hydrogen escape rate as a function of the mean temperature in the HD\,97658 b isotropic atmosphere. Lines in the lower part of the plot correspond to the 1$\sigma$ confidence level, lines in the upper part to the 3$\sigma$ level. Solid lines correspond to $R_{\mathrm{trans}}$ = 4\,$R_{\mathrm{pl}}$ and outflow bulk velocities $v_{\mathrm{trans}}$ = 1\,km\,s$^{-1}$ (blue), 10\,km\,s$^{-1}$ (green), and 20\,km\,s$^{-1}$ (red). Dashed lines correspond to $v_{\mathrm{trans}}$ = 10\,km\,s$^{-1}$ and transitions between the isotropic atmosphere and the exosphere $R_{\mathrm{trans}}$ = 2\,$R_{\mathrm{pl}}$ (blue), 4\,$R_{\mathrm{pl}}$ (green), and 6\,$R_{\mathrm{pl}}$ (red). The temperature is limited to the equilibrium temperature of the planet (T$_{eq}\sim$725\,K, \citealt{Knutson2014}).}
\label{chi2_contour_plot}
\end{minipage}
\end{figure*}

\begin{figure}
\centering
\includegraphics[trim=0.45cm 3.1cm 2.5cm 1cm,clip=true,width=\columnwidth]{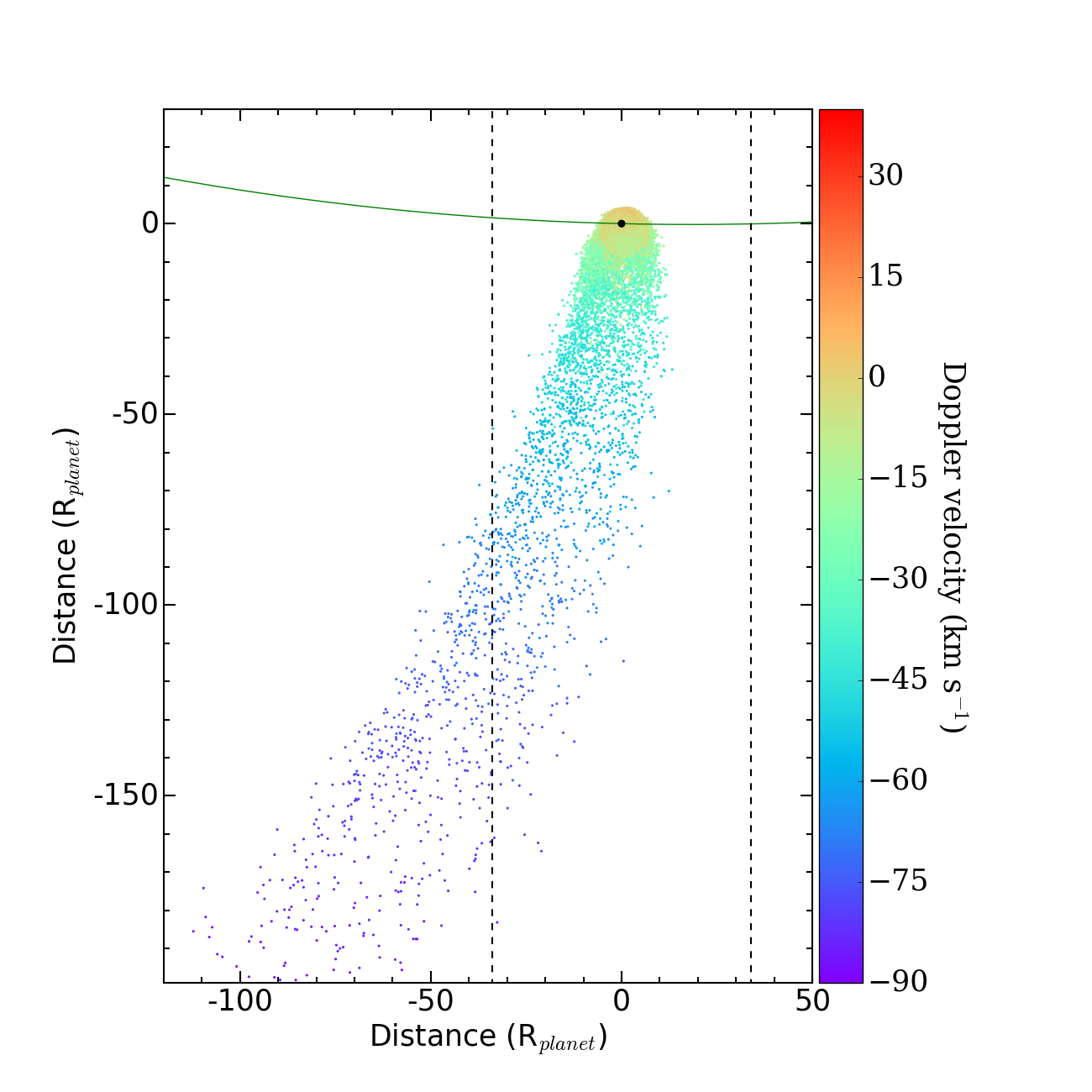}
\includegraphics[trim=0.45cm 0.5cm 2.5cm 3cm,clip=true,width=\columnwidth]{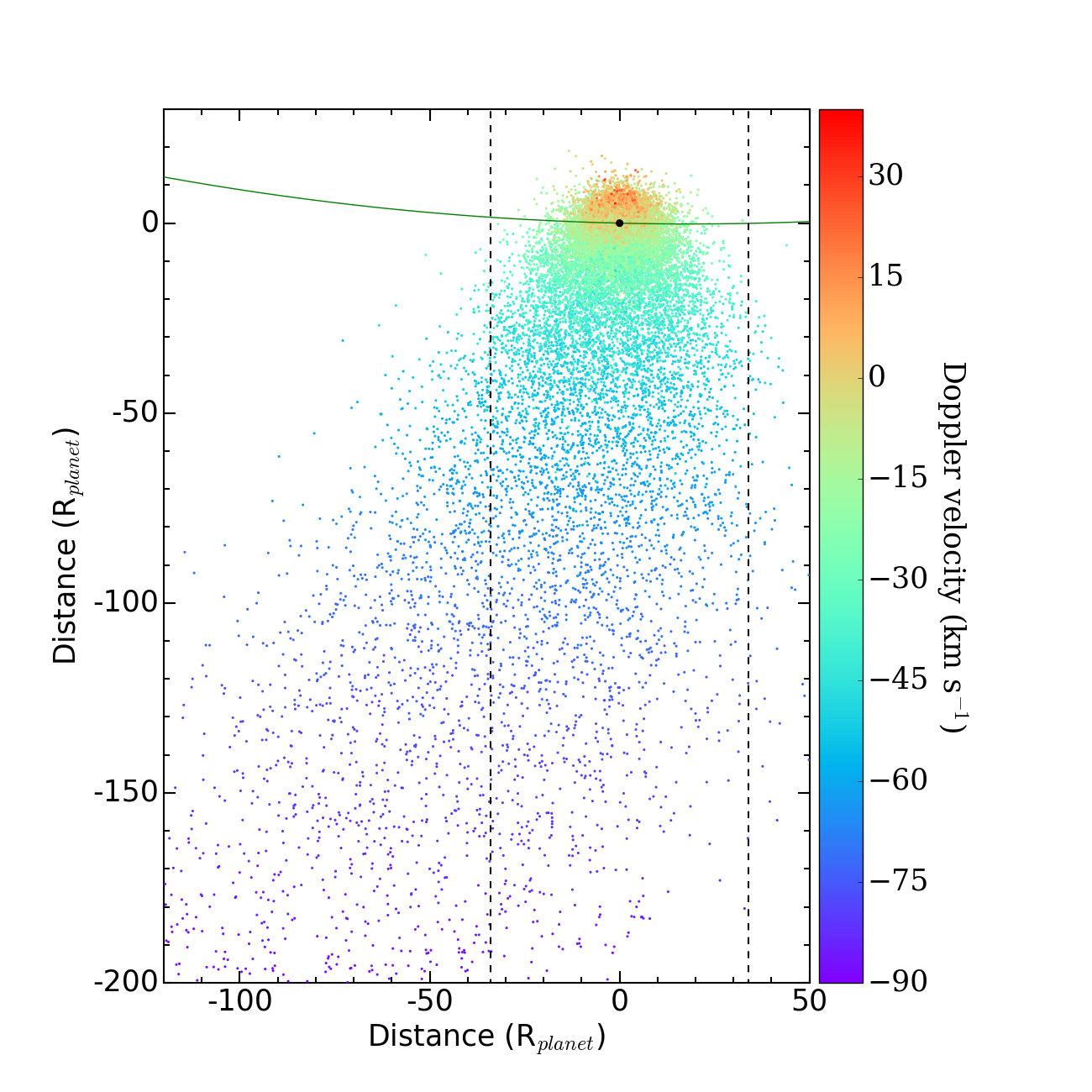}

\caption[]{Views of HD\,97658 b theoretical exosphere from the perpendicular to the orbital plane. Neutral hydrogen particles are colored as a function of their Doppler radial velocity according to the scale on the right axis. The planet is the small disk on the planetary orbit (green line). The dashed black lines limit the LOS toward Earth (the star is toward the top of the figure). Although the transit observations in the Lyman-$\alpha$ line are consistent with the absence of atmospheric escape, they still allow for the formation of a compact (\textit{upper panel}, $R_{\mathrm{trans}}$ = 3 $R_{\mathrm{pl}}$, $\dot{M}_{\mathrm{H^{0}}}$ = 2$\times$10$^{7}$\,g\,s$^{-1}$, $v_{\mathrm{trans}}$ = 1\,km\,s$^{-1}$, $T_{\mathrm{iso}}$ = 1000\,K) or low-density exosphere (\textit{lower panel}, $R_{\mathrm{trans}}$ = 6 $R_{\mathrm{pl}}$, $\dot{M}_{\mathrm{H^{0}}}$ = 3$\times$10$^{6}$\,g\,s$^{-1}$, $v_{\mathrm{trans}}$ = 10\,km\,s$^{-1}$, $T_{\mathrm{iso}}$ = 8000\,K). We note the comet-like shape of the exosphere, blown away by the strong radiation pressure, and its length that arises from the low photoionization rate.} 
\label{fig:ex_exo}
\end{figure}

The model atmosphere can be constrained in two ways by the observations. Neutral hydrogen in the isotropic atmosphere contributes mainly to absorption in the core of the Lyman-$\alpha$ line, which cannot be observed because of ISM absorption (Sect.~\ref{sect:longterm_var}). If the column density is high enough, the isotropic atmosphere can influence the observed spectra through the Lorentzian wings of its absorption profile (the so-called damping wings). However, this also requires a significant atmospheric expansion because damping wings can only absorb as much as the opaque surface of the atmosphere projected onto the stellar disk. It is highly unlikely that the atmosphere of a super-Earth like HD\,97658b would be dense within a large enough volume to produce detectable damping wings. Even if that were the case and the upper atmosphere extended up to the Roche lobe ($\sim$11.8\,R$_{p}$), its transit would still last for about the same duration as the optical transit ($\sim$3\,h). In contrast, the strong radiation pressure from the host star (Sect.~\ref{sec:intrinsic_values}) shapes the exosphere into an extended cometary tail trailing behind the planet (Fig.~\ref{fig:ex_exo}). We estimate from our simulations that the escaping neutral hydrogen can be accelerated by radiation pressure up to about 95\,km\,s$^{-1}$ away from the star, which implies that the exosphere can produce absorption mainly in the blue wing of the Lyman-$\alpha$ line after the optical transit. While the bulk motion of the gas in the exosphere is set by the combination of radiation pressure and stellar gravity, Fig.~\ref{fig:ex_exo} illustrates how its spatial extension in the transverse direction increases with increasing outflow velocity ($v_{\mathrm{trans}}$) and thermal dispersion ($T_{\mathrm{iso}}$). \\
Figure~\ref{chi2_contour_plot} shows upper limits on the escape rate of neutral hydrogen as a function of the mean temperature of the isotropic atmosphere and the outflow velocity. The transit observations of HD\,97658b are most consistent with the absence of an extended upper atmosphere, but they do not completely exclude a moderate atmospheric escape. With high values for the outflow temperature and bulk velocity, the escaping neutral hydrogen is diluted enough for the exosphere to be optically thin and its absorption roughly proportional to the local density. Since exospheric density scales at first order with the escape rate, the non-detection of an absorption signature from HD\,97658 b limits the escape rate roughly independently of the other model parameters for temperatures above $\sim$4000\,K (Fig.~\ref{chi2_contour_plot}). Conversely, the entire exosphere is more compact when neutral hydrogen escapes from low altitudes and/or with a lower outflow velocity and temperature (Fig.~\ref{fig:ex_exo}). In that case regions close to the planet become optically thick and their absorption increases more slowly with increasing escape rates, allowing for larger mass loss to be consistent with the observations (Fig.~\ref{chi2_contour_plot}). Finally, we note that the upper limit on the escape rate drops quickly when the temperature decreases below $\sim$1000\,K. This occurs because the density profile in the isotropic atmosphere is scaled by its value at the transition $R_\mathrm{trans}$, which depends on the escape rate. Since the density profile then increases steeply closer to the planet with low temperatures, even moderate escape rates can lead to high column density and the formation of strong damping wings at such temperatures, which are excluded by the observations. This effect is stronger when the size of the isotropic atmosphere increases as its larger projected area allows for larger damping wings.\\ 
\citet{Salz2016b} performed 1D, spherically symmetric simulations of HD\,97658b and predicted that the upper atmosphere of the planet should expand hydrodynamically, reaching velocities between $\sim$1 and 10\,km\,s$^{-1}$ and a mean temperature of about 3500\,K above 2\,R$_{p}$. In these conditions, our simulations set an upper limit of $\sim$10$^{8}$\,g\,s$^{-1}$ on the neutral hydrogen mass loss, which remains unchanged for larger temperature or stronger atmospheric expansion (Fig.~\ref{chi2_contour_plot}). Since it is unlikely that the upper atmosphere of HD\,97658 b could expand with lower temperatures \citealt{Salz2016b}, we consider that the upper limit on the neutral hydrogen mass loss from HD\,97658 b is 10$^{8}$\,g\,s$^{-1}$ at the 3$\sigma$ level.  \\

\subsection{Mass loss, heating efficiency, and neutral fraction}

Assuming that the upper atmosphere of HD\,97658 b is in the energy-limited regime (see, e.g., \citealt{Owen2016}), we calculated its total mass-loss rate as
\begin{equation}
\label{eq:H_esc_rate}
\dot{M}^{tot}= \eta \, (\frac{R_\mathrm{XUV}}{R_\mathrm{p}})^2 \, \frac{3 \, F_\mathrm{XUV}(\mathrm{sma})}{4 \, G \, \rho_\mathrm{p} \, K_{tide}}  
\end{equation}
with $\eta$ the fraction of the energy input that is available for atmospheric heating (e.g., \citealt{Ehrenreich_desert2011}, \citealt{Lammer2013}; \citealt{Shematovich2014}), $R_\mathrm{XUV}$ the mean altitude at which the XUV energy is absorbed in the upper atmosphere (e.g., \citealt{Lammer2003}), $F_\mathrm{XUV}(\mathrm{sma})$ the total XUV flux per unit area at the semi-major axis (sma) of HD\,97658 b, $\rho_\mathrm{p}$ the density of the planet, and $K_\mathrm{tide}$ a correction factor accounting for the contribution of tidal forces to the potential energy (\citealt{Erkaev2007}\footnote{Instead of using the Lagrangian points as the radius of the Roche lobe, we prefer to use the radius of the sphere with equivalent volume (\citealt{Eggleton1983}, \citealt{Ehrenreich2010}).}). Using our measurement for the XUV flux (835\,erg\,s$^{-1}$\,cm$^{-2}$, Table~\ref{tab:XEUV_flux}), we found $\dot{M}^{tot}$ = $\eta \, (\frac{R_\mathrm{XUV}}{R_\mathrm{p}})^2$\,2.8$\times$10$^{9}$\,g\,s$^{-1}$. Defining $\dot{M}_\mathrm{H^{0}}=f_\mathrm{H^{0}}\,\dot{M}^{tot}$, where $f_\mathrm{H^{0}}$ is the fraction of neutral hydrogen in the escaping gas and $\dot{M}_\mathrm{H^{0}}$ the upper limit on the escape rate of neutral hydrogen obtained with EVE, we can derive that $\eta\,(\frac{R_\mathrm{XUV}}{R_\mathrm{p}})^2\,f_\mathrm{H^{0}} < $3.6$\times$10$^{-2}$. This constraint hints at a low heating efficiency and/or low neutral hydrogen fraction in the upper atmosphere. Assuming $R_\mathrm{XUV}\,\sim\,R_\mathrm{p}$ and using theoretical estimations of $\eta$ between 10 and 30\% (e.g., \citealt{Lammer2013}; \citealt{Shematovich2014}; \citealt{Owen2016}; \citealt{Salz2016a}), the upper limit on the neutral hydrogen fraction would range between about 12 and 36\%.\\
An independent constraint can be derived from the total mass-loss rate $\dot{M}^{tot}$ = 3$\times$10$^{9}$ g\,s$^{-1}$ predicted by \citet{Salz2016b}. They used an XUV irradiation of 955\,erg\,s$^{-1}$\,cm$^{-2}$, which is consistent with the value we measure. However, the comparison between their $\dot{M}^{tot}$ and our 3$\sigma$ upper limit on the neutral hydrogen escape rate $\dot{M}_\mathrm{H^{0}}$ yields $f_\mathrm{H^{0}} \approxinf$3\%. This value is surprising considering the low irradiation and photoionization rate derived from our observations (Sect.~\ref{sec:photoion}), and is inconsistent with the high neutral hydrogen fraction ($>$50\%) obtained by \citet{Salz2016b}. This suggests either that the upper atmosphere of HD\,97658 b does not expand hydrodynamically, or that its thermospheric and mass-loss properties are much different from the ones predicted by \citet{Salz2016b}.\\

\section{Discussion}
\label{sec:discuss}

Despite appearing quiet in the optical and NIR (\citealt{Vangrootel2014}), HD\,97658 has detectable high-energy emission. It is a weak and soft X-ray source with a variable upper chromosphere and corona. Our HST and XMM-Newton/Chandra observations of HD\,97658 at five different epochs reveal signs of both short-term and long-term variability in the Lyman-$\alpha$ and X-ray emission of this star. We were nonetheless able to  determine a stable, average reference for the intrinsic Lyman-$\alpha$ line and XUV spectrum of the star, refining the properties derived by \citet{Youngblood2016}. These estimates will be useful in determining the impact of the incident high-energy radiation on the atmospheric heating and chemistry for this super-Earth, and more generally in understanding the conditions that lead to evaporation. The HD\,97658 system might be crucial in understanding the stability of the atmospheres of  small planets in terms of atmospheric composition, planetary density, and irradiation. Indeed, HD\,9758b is a super-Earth more than twice as dense (3.9\,g\,cm$^{-3}$) as the evaporating warm Neptune GJ\,436 b (1.6\,g\,cm$^{-3}$). Furthermore, HD\,97658 b orbits at 0.08\,au from its K dwarf host star and we measure an XUV energy input of 835$\pm$160\,erg\,s$^{-1}$\,cm$^{-2}$ (Sect.~\ref{sec:XUV_irr}), which is about three times lower than the irradiation of GJ\,436 b (at 0.027\,au from its M dwarf host; \citealt{Ehrenreich2015}, \citealt{Bourrier2016}). The upper atmosphere of HD\,97658b is also subjected to a photoionization rate of 4.8$\stackrel{+1.6}{_{-1.4}}\times$10$^{-5}$\,s$^{-1}$, which is more than six times lower than for the evaporating hot Jupiter HD\,189733 b (at 0.031\,au from its K dwarf host; \citealt{Bourrier_lecav2013}).\\

Our analysis of three Lyman-$\alpha$ transits of the super-Earth HD\,97658b at independent epochs yields no signature arising from an extended atmosphere of neutral hydrogen. This suggests that the thermosphere of HD\,97658b is not hydrodynamically expanding, and in any case that the escape of neutral hydrogen is below 10$^{8}$\,g\,s$^{-1}$ at 3$\sigma$. This might be linked to the low XUV stellar irradiation, to an inefficient conversion of the stellar energy input, and/or to a low hydrogen content in the upper atmosphere (possibly linked to a helium enrichment, \citealt{Hu2015}, or to an enhanced mass loss from the irradiation and stellar wind of the young host star). A more exotic possibility would be that HD\,97658b is in a saturation regime with the wind from its host star, as defined by \citet{Bourrier_lecav2013}, where the stellar wind is so dense that it interacts with all neutral hydrogen atoms escaping the upper atmosphere. Provided that the stellar wind protons move faster than about 300\,km\,s$^{-1}$, all hydrogen atoms escaping HD\,97658b would then be ionized through charge-exchange, while the high-velocity neutralized protons would contribute to Lyman-$\alpha$ absorption too far from the line core to be detectable (\citealt{Bourrier2016}).\\

The methodology used in this study is the same as led to the non-detection of Lyman-$\alpha$ signatures around the super-Earth 55\,Cnc e and to the detection of hydrogen exospheres around the giant planets HD\,209458 b (\citealt{VM2003}), HD\,189733 b (\citealt{Lecav2010}; \citealt{Lecav2012}), 55\,Cnc b (\citealt{Ehrenreich2012}), and GJ\,436b (\citealt{Ehrenreich2015}). There is a clear distinction between the stellar activity of HD\,97658 b, revealed as short- and long-term variations in the core of the Lyman-$\alpha$ line, and the transit of a neutral hydrogen exosphere. The latter occurs around and after the time of the optical transit, with the depth and spectral range of the absorption evolving coherently over time. Confirmed signatures of hydrogen escape have been detected within specific wavelength ranges in the blue wing of the Lyman-$\alpha$ line, which can be  explained well by the physical mechanisms acting on the upper planetary atmosphere. In this frame our study of HD\,97658 b can be seen as a control experiment; the results we obtained confirm that detections of atmospheric escape are not biased by our methodology, and underline its strength and reliability. Future observations of this system in other FUV lines will be needed to search for escape signatures of heavier species than hydrogen and to further constrain the nature of this super-Earth.\\

\begin{acknowledgements}
We thank the referee, Jeffrey Linsky, for his comments that helped improve the science and clarity of this paper. We warmly thank Diana Dragomir for her validation of the HD\,97658b transit ephemeris. This work is based on observations made with the NASA/ESA Hubble Space Telescope, obtained at the Space Telescope Science Institute (STScI), which is operated by the Association of Universities for Research in Astronomy, Inc., under NASA contract NAS 5-26555. The authors thank the STScI for extending the proprietary period of program 13820, allowing a careful analysis of the full dataset. This work has been carried out in the frame of the National Centre for Competence in Research ``PlanetS'' supported by the Swiss National Science Foundation (SNSF). V.B. and D.E. acknowledge the financial support of the SNSF. A.L.E acknowledges financial support from the Centre National d'\'{E}tudes Spatiales (CNES). The authors acknowledge the support of the French Agence Nationale de la Recherche (ANR), under program ANR-12-BS05-0012 ``Exo-Atmos''. 
\end{acknowledgements}

\bibliographystyle{aa} 
\bibliography{biblio} 

\end{document}